\documentclass[%
 reprint,
 amsmath,amssymb,superscriptaddress,
 aps,showkeys
]{revtex4-2}

\usepackage{graphicx}
\usepackage[caption=false]{subfig} 
\usepackage{dcolumn}
\usepackage{bm}
\usepackage{physics}
\usepackage{mathtools,mathrsfs,amsfonts,dsfont}
\usepackage{calrsfs}
\usepackage{scalerel}
\usepackage{ulem}
\usepackage{float}
\usepackage{verbatim}
\usepackage{lineno}

\newcommand{\tpsi}{\skew{3}{\tilde}{\psi}}
\newcommand{\PP}{\mathbb{P}}
\newcommand{\sm}{\kern0.1em}

\newcommand{\syst}[1]{\left\{ \begin{aligned} #1 \end{aligned}  \right.}

\usepackage{hyperref, xcolor}
\usepackage[nameinlink,capitalize]{cleveref}
\hypersetup{
	colorlinks = true,
	linkbordercolor = {white},
	linkcolor={blue},
	citecolor={blue},
	urlcolor={blue}
}

\usepackage{cleveref}
\usepackage{orcidlink}
\usepackage{footmisc}

\begin{document}

\preprint{APS/123-QED}

\title{Can the double-slit experiment distinguish between quantum interpretations?}

\author{Ali Ayatollah Rafsanjani\,\orcidlink{0000-0002-3717-718X}}
	\email{aliayat@physics.sharif.edu}
	\affiliation{Department of Physics, Sharif University of Technology, Tehran, Iran}
	\affiliation{School of Physics, Institute for Research in Fundamental Sciences (IPM), Tehran, Iran}

\author{MohammadJavad Kazemi\,\orcidlink{0000-0001-7709-7793}}
	\affiliation{Department of Physics, Faculty of Science, University of Qom , Qom, Iran}
	
	\author{Alireza Bahrampour}
	\affiliation{Department of Physics, Sharif University of Technology, Tehran, Iran}
	\author{Mehdi Golshani}
	\affiliation{School of Physics, Institute for Research in Fundamental Sciences (IPM), Tehran, Iran}


\begin{abstract}
Despite the astonishing successes of quantum mechanics, due to some fundamental problems such as the measurement problem and quantum arrival time problem, the predictions of the theory are in some cases not quite clear and unique. Especially, there are various predictions for the joint spatiotemporal distribution of particle detection events on a screen, which are derived from different formulations and interpretations of the quantum theory. Although the differences are typically small, our studies show that these predictions can be experimentally distinguished by an unconventional double-slit configuration, which is realizable using present-day single-atom interferometry. This experiment would enrich our understanding of the foundations of quantum mechanics.
\end{abstract}
 
\maketitle


\section{Introduction}\label{intro}
In textbook quantum theory, time is a parameter, not a self-adjoint operator \footnote{As shown by Pauli \cite{Pauli1958Encyclopedia}, if the Hamiltonian spectrum is discrete or has a lower bound, then there is no self-adjoint time operator canonically conjugate to the Hamiltonian.}, hence there is no agreed-upon way to compute the temporal probability distribution of events from the first principles (i.e. the Born rule) 
\footnote{After decades of theoretical debate on this topic \cite{muga2000arrival,muga2007time}, it seems that the only agreed-upon opinion is that, in general cases, there is no agreed-upon opinion for computing temporal distributions. On the one hand, some authors argue that, despite Pauli's objection \cite{galapon2002pauli,leon2017pauli}, introducing a proper time operator is possible in various way \cite{art:Grot-Rovelli-Tate,kijowski1999comment,galapon2004confined,galapon2005transition,galapon2012only,flores2022relativistic}. On the other hand, some other authors argue that, despite quantum Zeno paradox \cite{misra1977zeno,porras2014quantum}, the temporal distribution of events can be treated in standard interpretation without directly using of a self-adjoint time operator and the result is the same as the result of Bohmian interpretation \cite{Tumulka2022Distribution,tumulka2022absorbing,Tumulka2022Detection,jurman2021time,juric2022arrival,juric2022passive}. While some other authors discuss the differences in the results of the various approaches \cite{leavens2002standard,egusquiza2003comment,leavens2005reply,nitta2008time,vona2013does,sombillo2016particle,Zimmermann2016,
anastopoulos2017time,anastopoulos2019decays,Das2019Arrival,das2021questioning,das2021times,roncallo2023does}
\label{foot2}}.
Nonetheless, since clocks exist and time measurements are routinely performed in quantum experiments \cite{Zimmermann2016,Kataoka2016}, a complete quantum theory must be able to predict the temporal statistics of detection events. For example, in the famous double slit experiment, each particle is detected at a \textit{random time} as same as at a \textit{random position} on the detection screen \cite{kolenderski2014time,Frabboni2012,kurtsiefer1997measurement,nitta2008time, Das2019Exotic,das2022double}. Therefore, one can ask: What is the position-time \textit{joint} probability density $\mathbb{P}(\vb x, t)$
on the screen? Although this question is very old \cite{Allcock1969,Kijowski1974,Werner_Screen,mielnik1994screen,Marchewka2001Survival}, it is still open \cite{vona2013does,Maccone2020,Dias2017,das2021times,art:Das,kazemi2022detection,Rafsanjani2023Non}. In fact, the ambiguity in the arrival time distribution even prevents a clear prediction of cumulative arrival position distribution, $\int \mathbb{P}(\vb x, t) dt$, which is typically measured in a non-time-resolved double-slit experiment \footnote{Note that, the Heisenberg position operator describes position measurement at a specific time, not position measurements at random times \cite{DGZ2004,durr2004exit}. In other words, $|\psi_t(\vb x)|^2$ is just the \textit{conditional} position probability density $\mathbb{P}(\vb x|t)$ \cite{Maccone2020,Dias2017,Terno2014,sombillo2016particle}, not the position-time \textit{joint} probability density $\mathbb{P}(\vb x,t)$ \cite{kijowski1999comment,daumer1997,das2022double}\label{foot2}}.

Nonetheless, usual experiments are performed in the far-field (or scattering) regime, where a semiclassical analysis is often sufficient \cite{vona2013does,Note2}. In this analysis, it is assumed that particles move along classical trajectories, and the arrival time distribution is computed using the quantum momentum distribution
\cite{Das2019Exotic,Shucker1980, wolf2000ion}. However, because of the \textit{quantum backflow} effect \cite{bracken1994probability}, even in free space, the quantum mechanical time evolution of position probability density is not consistent with the underlying uniform motion assumption, especially in near-field interference phenomena \cite{Hofmann2017}. In fact, due to recent progress in the ultra-fast detectors technology (e.g. see \cite{Korzh2020,Steinhauer2021,Azzouz12012,Rosticher2010}), it will be soon possible to investigate the near-field regime, where the semiclassical approximation breaks down and a deeper analysis would be demanded \cite{vona2013does,lgado2006}.

To remedy this problem, based on various interpretations and formulations of quantum theory, several attempts have been made to introduce a suitable arrival time distribution. On the one hand, according to the (generalized) standard canonical interpretation, the arrival distribution is considered as a generalized observable, which is described by a positive-operator-valued measure (POVM), satisfying some required symmetries \cite{Kijowski1974,Werner_Screen,Hegerfeldt_2010,Hegerfeldt2010Manufacturing}. On the other hand, in the realistic-trajectory-based formulations of quantum theory, such as the Bohmian mechanics \cite{PhysRev.85.166}, Nelson stochastic mechanics \cite{PhysRev.150.1079}, and many interacting worlds interpretation \cite{PhysRevX.4.041013}, the arrival time distribution could be obtained from particles trajectories \cite{Leavens_BohmianTOA,nitta2008time,Das2019Arrival,kazemi2022detection,Rafsanjani2023Non}. Moreover, in other approaches, the arrival time distribution is computed via phenomenological modeling of the detection process, such as the (generalized) path integral formalism in the presence of an absorbing boundary \cite{Marchewka1998Feynman,Marchewka2000Path,Marchewka2001Survival,Marchewka2002}, Schrödinger equation with complex potential or absorbing boundary \cite{werner1987arrival,Tumulka2022Distribution,Tumulka2022Detection,Dubey2021Quantum,tumulka2022absorbing}, and so on \cite{juric2022arrival,jurman2021time,juric2022passive,roncallo2023does,damborenea2002measurement,muga1995time,Halliwell2008Path}.

In principle, the results of these approaches are different. However, in most of the experimental situations, the differences are typically slight, and so far as we know, in the situation where differences are significant, none of the proposals have been backed up by experiments in a strict manner \cite{Das2019Arrival,Das2019Exotic}.
An experiment that can probe these differences would undoubtedly enrich our understanding of the foundations of quantum mechanics. The purpose of the present paper is to make it evident, via numerical simulations, that the famous two-slit experiment could be utilized to investigate these differences if we simply use a horizontal screen instead of a vertical one: see Fig.\,\ref{SchematicSetup}. 
Using current laser cooling and magneto-optical trapping technologies, this type of experiment can be realized by Bose-Einstein condensates, as a controllable source of coherent matter waves \cite{andrews1997observation,Shin2004Atom,Cronin2009Optics}. 
Moreover, our numerical study shows that the required space-time resolution in particle detection is achievable using fast single-atom detectors, such as the recent delay-line detectors described in \cite{Keller2014Bose,khakimov2016ghost} or the detector used in \cite{kurtsiefer19962A,kurtsiefer1997measurement}.

The structure of this paper is as follows: In Section \ref{sec_Methods}, we study the main proposed \textit{intrinsic} arrival distributions. Then, in section \ref{sec3} we compare them in the double-slit setup with vertical and horizontal screens and in different detection schemes. In Section \ref{screen_back_effect}, we study the screen back-effect, and we summarize in section \ref{summary}.

\section{``Intrinsic" arrival distributions}\label{sec_Methods}
In this section, we first review the semiclassical approximation and then scrutinize two main proposed intrinsic arrival time distributions \cite{Das2019Arrival,das2021times} and their associated screen observables. In these approaches, the effect of the detector's presence on the wave function evolution, before particle detection, is not considered. We discuss this effect in section \ref{screen_back_effect}. 
\subsection{Semiclassical approximation}
As mentioned, in the experiments in which the detectors are placed far away from the support of the initial wave function (i.e. far-field regime), the semiclassical arrival time distribution is routinely used to the description of the particle time-of-flight \cite{gliserin2016high,wolf2000ion,kurtsiefer1995time,copley1993neutron,kothe2013time}. In this approximation, it is assumed that particles move classically between the preparation and measurement. In this approach, the arrival time randomness is understood as a result of the uncertainty of momentum, and so the arrival time distribution is obtained from momentum distribution \cite{vona2013does,vona2015role,art:Das,Das2019Arrival}. In the one-dimensional case, the classical arrival time is given by
\begin{equation}
  t =m(L-x_0)/p_0,
 	\label{SC0}
\end{equation}
which is applicable for a freely moving particle of mass $m$ that at the initial $t\!=\!0$ had position $x_0$ and momentum $p_0$ arriving at a distant point $L$ on a line. Hence, for a particle with the momentum wave fuction $\tpsi_0(p)$, assuming $\Delta x_0\!\ll \! |L-\langle x\rangle_0|$,  the semiclassical arrival time distribution reads \cite{vona2015role}
\begin{equation}
  \Pi_{\text {SC}}(t|x\!=\!L) =\frac{mL}{t^2} |\tpsi_0(mL/t)|^2. 
 	\label{SC1}
\end{equation}

This analysis could be generalized in three-dimensional space. Then,
the distribution of arrival time at a screen surface $ \mathbb{S}$ is given by \cite{Das2019Arrival}
\begin{equation}
	\Pi_{\text {SC}}(t | \vb x \! \in \! \mathbb{S}) = \frac{m^3}{t^4}\int_{\mathbb{S}} |\tpsi_0(m\vb{x}/t)|^2 \, \vb{x} \cdot d \vb{S},
	\label{SC2}
\end{equation}
where the  $d\vb{S}$ is the surface element directed outward. The other main distribution that should be demanded is the joint position-time probability distribution on the screen, which is also called "screen observable" \cite{Werner_Screen}. 
Using the conditional probability definition, the joint probability of finding the particle in $dS$ and in a time interval $[t, t \!+\! dt]$ could be written as
$
\PP(\vb x,t | \vb x \! \in \! \mathbb{S})dSdt=\left[\Pi(t | \vb x \! \in \! \mathbb{S})dt\right] \times \left[\PP(\vb x | \vb x \! \in \! \mathbb{S},t) dS\right]. 
$
In this regard, one can use the fact that $\psi_t(\vb{x})$ is the state of the system, conditioned on the time being $t$ in the Schrödinger picture. This implies that $|\psi_t(\vb x)|^2$ refers to the position probability density conditioned at a specific time $t$ \cite{Maccone2020,Dias2017,arce2012unification}. Therefore, in the semiclassical approximation, the joint spatiotemporal probability density reads as
\begin{equation}
    \PP_{\text {SC}}(\vb x,t | \vb x \! \in \! \mathbb{S}) =N_{\text {SC}}\Pi_{SC}(t | \vb x \! \in \! \mathbb{S}) \left| \psi_t(\vb x) \right|^2
 	\label{SC3}
\end{equation}
in which  $N_{\text {SC}} \!\equiv\! 1/\int_{\mathbb{S}} dS \left| \psi_t(\vb x ) \right|^2$ is the normalization constant,
and $dS \!=\! \vb n \cdot d\vb S$, where ${\vb n}$ is the outward unit normal vector at $x\!\in\!\mathbb{S}$. 

\subsection{``Standard" approach}
The first attempts to investigate the arrival time problem, based on the standard rules of quantum theory, were made at the beginning of the 1960s by Aharonov and Bohm \cite{aharonov1961time}, and also Paul \cite{paul1962quantenmechanische}. This approach starts with a symmetric quantization of classical arrival time expression  \eqref{SC0}, as follows \cite{muga2000arrival}:
\begin{equation}
     \hat{t}_{AB}=mL \, \hat{p}^{\,-1}-\frac{m}{2}\,(\hat{p}^{\,-1}\,\hat{x}\,+\,\hat{x}\,\hat{p}^{\,-1}),   
     \label{ABquant}
\end{equation}
where $\hat{x}$ and $\hat{p} \!=\! -i \, \partial / \partial x$  are the usual position and momentum operators, respectively, and $\hat{t}_{AB}$ is called  the Aharonov-Bohm time operator. This operator satisfies the canonical commutation relation with the free Hamiltonian operator, $[\hat{t}_{AB}, \hat{p}^2/2m]\!=\!i\hbar$, which has been used to establish the energy-time uncertainty relation \cite{giannitrapani1997positive,art:Grot-Rovelli-Tate}. However, although $\hat t_{AB}$ is Hermitian (or symmetric in mathematics literature), it is not a self-adjoint operator \cite{egusquiza1999free}---a fact that is in agreement with Pauli's theorem \cite{Note1}. The origin of this non-self-adjointness can be understood as a result of the singularity at $p\!=\!0$ in the momentum representation, $\hat{t}_{AB}\to (i\hbar m/2)(p^{-2}-2p^{-1}\partial_p)$ 
\cite{egusquiza1999free}.
 Nevertheless, although the (generalized) eigenfunctions of $\hat{t}_{AB}$ are not orthogonal, they constitute an over-complete set and  provide a POVM, which are used to define the arrival-time distribution as follows \cite{giannitrapani1997positive,egusquiza1999free}:
\begin{equation}
   \!\! \Pi_{\text{STD}}(t|x\!=\!L) \!=\! \frac{1}{2\sm\pi \hbar}\!\sum_{\alpha\sm=\sm\pm}\! \left| \int_{-\infty}^{\infty}\!\!\!\!\!\!dp\, \theta(\alpha p)\,\sqrt{\frac{|p|}{m}}  \tpsi_t(p) e^{\frac{i}{\hbar}Lp}\right|^2\!\!\!\!,
    \label{ABdis}
\end{equation}
where $\theta(\cdot)$ is Heaviside's step function and $\tpsi_t(p)$ is the wave function in the momentum representation which could be obtained from the initial wave function $\tpsi_0(p)$, as
$
\tpsi_t(p)= \tpsi_0(p)\exp[-\,it p^2/2m\hbar].
$ 
The distribution $\Pi_{\text{STD}}$ and its generalization in the presence of interaction potential have been referred to as the "standard arrival-time distribution" by some authors \cite{muga2007time,egusquiza2003comment,leavens2005spatial,leavens2007peculiar,das2021times}.
In fact, Grot, Rovelli, and Tate treated the singularity of \eqref{ABquant} by symmetric regularization and obtained equation \eqref{ABdis} via the standard Born rule \cite{art:Grot-Rovelli-Tate}.
The generalizations of equations \eqref{ABquant} and \eqref{ABdis} in the presence of interaction potential have been investigated in various works \cite{galapon2004shouldn,galapon2004confined,galapon2005transition,galapon2009theory,flores2019quantum,galapon2012only, Hegerfeldt2010Manufacturing,das2021times}. 
Using these developments, it has been shown that the non-self-adjointness of the free arrival time operator can also be lifted by spatial confinement \cite{galapon2002self,galapon2004confined}, and the above arrival time distribution could be derived from the limit of the arrival time distribution in a confining box as the length of the box increases to infinity \cite{galapon2005transition}. Furthermore, recently, the  distribution (\ref{ABdis}) is derived from a space-time-symmetric extension of non-relativistic quantum mechanics \cite{dias2017space}. 

The three-dimensional generalization of \eqref{ABdis} is derived by Kijowski's \cite{Kijowski1974} via an axiomatic approach. The assumed axioms are implied by the principle of the probability theory,  the mathematical structure of standard  quantum mechanics, and the Galilei invariance \cite{kijowski1999comment}. Based on these axioms, Kijowski constructed the following arrival time distribution for a free particle that passes through a two-dimensional plane $\mathbb{S}$ as
\begin{eqnarray}
   \!\!\!\!\!\!\!\!  \Pi_{\text{STD}}(t | \vb{x}\in  \mathbb{S})\!\!\!\!\!\!&&=\frac{1}{2\sm\pi \hbar\sm}\!\sum_{\alpha\sm=\sm\pm}\int_{\mathbb{R}^2}\!\!d^2 \vb p_{\|} \nonumber \\
     &&\!\!\!\!\!\!\!\!\!\!\!\!\!\!\! \times\,\left| \int_{-\infty}^{\infty}\!\!\!d p_{\bot}\,\theta(\alpha \vb p \! \cdot \! \vb n)\sqrt{\frac{|\vb p_{\bot}|}{m}}~\tpsi_t(\vb{p})e^{\frac{i}{\hbar} \vb x  \cdot \vb p_{\bot}}\sm\right|^2\!\!,
    \label{3Dkij}
\end{eqnarray}
where $\vb{p}_{\bot}\!\equiv\!\!(\vb p  \cdot   \vb n)\vb n$ and $\vb{p}_{\|}\!\equiv\!\vb p-\vb{p}_{\bot}$ are perpendicular and parallel components of $\vb p$ relative to $ \mathbb{S}$ respectively, and $\vb n$ is the outward normal of plane $ \mathbb{S}$. In fact, he first proves the above expression for the wave functions whose supports lie in the positive (or  negative) amounts of ${p}_{\bot}$. Then he \textit{uniquely} derives the following self-adjoint variant of the (three-dimensional version of)  Aharonov-Bohm arrival time operator, by demanding that the time operator be self-adjoint and leads to \eqref{3Dkij} for these special cases via the Born rule \cite{Kijowski1974,kijowski1999comment}:
\begin{equation}
\hat{t}_{L}=\text{sgn}(\hat p_{\bot}) \left[ mL\hat p_{\bot}^{-1}-\frac{m}{2}(\hat p_{\bot}^{-1}\hat x_{\bot}+\hat x_{\bot}\hat p_{\bot}^{-1})\right],
\end{equation}
where $\hat x_{\bot}\equiv\hat{\vb x}\! \cdot \!\vb n$ and $L$ ($\equiv\vb x \! \cdot \!\vb n$) represent the distance between the detection surface and the origin \footnote{The presence of $\text{sgn}(\hat p_{\bot})$ operator ensures the self-adjointness of this time operator, however, leads to a modified commutation relation, i.e. $[\hat{t}_K,\hat{H}]=i\hbar\ \text{sgn}(\hat p_{\bot})$. }. Finally, for an arbitrary wave function, the equation \eqref{3Dkij}  could be derived from this self-adjoint operator.
Moreover, considering this time operator, besides the components of the position operator in the detection plane, $\hat{\vb x}_{\|}\equiv\hat{\vb x}-(\hat{\vb x}  \cdot  \vb n)\vb n$, Kijowski obtains the following expression as the joint position-time distribution on the detection screen  via the Born rule \cite{kijowski1999comment}:
\begin{equation}
    \PP_{\text{STD}} (\vb x,t| \vb{x}\! \in \! \mathbb{S})= \sum_{\alpha=\pm}|\psi^\alpha_\mathbb{S}(\vb x,t)|^2,
    \label{STDdis}
\end{equation}
in which $\psi^\pm_\mathbb{S}(\vb x,t)$ is the wave function on the basics of joint eigenstates of the operators $\hat{t}_L$ and $\hat{\vb x}_{\|}$. Explicitly
\begin{equation}
   \!\!\!\!\! \psi^\pm_\mathbb{S}(\vb x,t) =\frac{1}{(2\pi \hbar)^{3/2}}  \int d^3 \vb p \, \theta(\pm\vb p \! \cdot \! \vb n)\,\sqrt{\frac{|\vb p_{\bot}|}{m}} \tpsi_t(\vb p)  e^{\frac{i}{\hbar}\vb x  \cdot \vb p }.
   \label{psiSpm}
\end{equation}
Note that,  the arrival time distribution \eqref{3Dkij} could be reproduced by taking the integral of \eqref{STDdis} over the whole of the screen plane.
The joint space-time probability distribution \eqref{STDdis}, and its generalization for the particles with arbitary spin, have been also derived by Werner in another axiomatic manner \cite{Werner_Screen}. Moreover, it is easy to see that the results \eqref{3Dkij} and \eqref{STDdis} can be obtained from a regularized version of the (three-dimensional generalization of) Aharonov-Bohm time operator, which is the same as the procedure used by Grot, Rovelli and Tate in one-dimensional cases \cite{art:Grot-Rovelli-Tate}. However, some paradoxical behaviors have been raised about this distribution which we discuss in appendix \ref{appendixB}.
  
\subsection{Quantum flux and Bohmian approach}\label{subsec_QF}
Inspiring by classical intuition, another proper candidate for screen observables is the perpendicular component of the quantum probability current to the screen surface, $\vb{J}(\vb x,t)\! \cdot \!\vb{n}$, where
\begin{equation}
     \vb{J}(\vb x,t)=-\frac{\hbar}{m}\Im\left[\psi_t^*(\vb x)\grad\psi_t(\vb x)\right],
\end{equation}
and $\vb{n}$ is the outward normal to the screen $\mathbb{S}$. This proposal is applicable for a particle in a generic external potential and a generic screen surface, not necessarily an infinite plane. There are several attempts to derive this proposal in various approaches, such as Bohmian mechanics for the scattering case in \cite{daumer1997}, decoherent histories approach in \cite{halliwell2009quantum} as an approximation, or in \cite{boonchui2013arrival} as an exact formula using the concept of extended probabilities, and so on \cite{damborenea2002measurement,muga1995time,Hannstein2005}.
However, even if the wave function contains only momentum in the same direction as $\vb{n}$, the $\vb{J}(\vb x,t)\! \cdot \!\vb{n}$ could be negative due to the \textit{backflow} effect \cite{bracken1994probability}. This property is incompatible with the standard notion of probability.

 Nevertheless,  this problem could be treated from the Bohmian point of view: Using Bohmian trajectories, it can be shown that the positive and negative values of $\vb{J}(\vb x,t)\!\cdot \!\vb{n}$ correspond to the particles that reach the point $\vb{x}$ at $\mathbb{S}$ in the same direction of $\vb{n}$ or the opposite direction of it, respectively \cite{leavens1993arrival,PhysRevA.51.2748}. In this regard,  through the Bohmian mechanics in one-dimension, Leavens demonstrates that the time distribution of arrival to $x\!=\!L$ from both sides could be obtained from the absolute form of probability flux as \cite{LEAVENS1998795,Leavens_BohmianTOA}
\begin{equation}
   \Pi_{\text{QF}}(t|x\!=\!L) = \frac{|J(L,t)|}{\int dt \, |J(L,t)|},
       \label{QF1D}
\end{equation}
which is free from the aforementioned problem. Furthermore, recently Juric and Nikolic have treated this problem from a different point of view \cite{juric2022passive}. In the Juric-Nikolic analysis, the negative fluxes are interpreted as zero arrival probability density, which originates from a physical insight that in this case, the particle departs, rather than arrives.

\begin{figure}[H]
    \centering
    \includegraphics[width = 8.6cm]{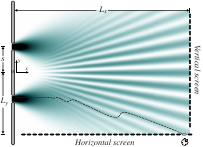}
     \caption{\textbf{Schematic double-slit experiment setup.} The center of two slits is considered as the coordinate origin and the distance between the two slits is $2s$. The vertical and horizontal screens are placed at $x\!=\!L_x$ and $y\!=\!L_y$, respectively. The dashed black line shows a typical Bohmian trajectory---with recursive movements---that arrives at the horizontal screen. A suitable single-particle detector, in addition to particle arrival position, can record the arrival time using a proper clock.}
    \label{SchematicSetup}
\end{figure}

\begin{figure}[h]
    \centering
    \includegraphics[width = 8.3 cm]{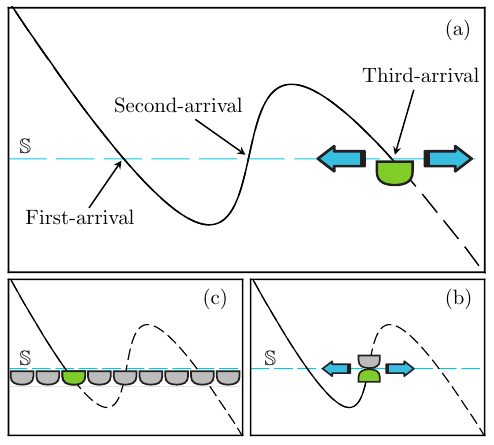}
   \caption{\textbf{Different schemes of particle detection on the screen surface $\mathbb S$.} In the Bohmian point of view, particles could have a recursive motion on surface $\mathbb S$ and cross it more than once (e.g. see the plotted trajectory in Fig.\,\ref{SchematicSetup}). Assuming different detector types, one can probe variant possible observables on the screen. In panel (a) a conceivable particle trajectory is depicted, which crosses $\mathbb S$ three times. In this panel, a movable point-like detector is placed on $\mathbb S$, which can survey the whole screen and detect particles that arrive only from one side, while in panel (b) a two-sided point detector is placed on $\mathbb S$, which can move along it and detect particles that arrive from up and down. In addition, one can assume there is (c) an array of side-by-side detectors covering the entire screen surface $\mathbb S$. The last configuration blocks the trajectory and does not allow the crossed particle to return. In this scheme, we only detect \textit{first-arrivals} from one side.}
    \label{DetectionSchemes}
\end{figure}

The three-dimensional justification of $\vb{J}(\vb x,t)\! \cdot \!\vb{n}$ as an operational formulation of the arrival time model has been made in \cite{Hannstein2005}. Also, the generalization of \eqref{QF1D} for arrival to the surface $\mathbb{S}$ is given by  \cite{vona2013does,das2021times,nitta2008time,AliPhysRevA}
\begin{equation}
   \Pi_{\text{QF}}(t|\vb x\!\in\!\mathbb{S}) = \frac{\int_{\mathbb{S}} dS |\vb J(\vb x,t)\!\cdot\!\vb n|}{\int dt \int_{\mathbb{S}} dS  |\vb J(\vb x,t)\! \cdot \!\vb n|},
    \label{QF3D}
\end{equation}
with $dS\!=\!\vb n   \cdot  d\vb S$ the magnitude of the surface element $d \vb S$ which is directed outward at $\vb x\!\in\!\mathbb S$. To illustrate \eqref{QF3D} and to generalize it to the case of joint arrival distribution, we can use the Bohmian point of view. In this theory, each particle has a specific trajectory, depending on the initial position, and so the rate of passing particles through an area element $d\vb S$ centered at $\vb x\!\in\!\mathbb{S}$, in the time interval between $t$ and $t+dt$, is proportional to $ \rho_t(\vb x) |\vb v (\vb x,t)  \cdot  d\vb{S}|dt$, where $\vb{v}(\vb x,t)\!=\! \vb{J}(\vb x , t)/|\psi_t(\vb x)|^2$ is the Bohmian velocity of the particle. Hence,  using quantum equilibrium condition \cite{durr1992quantum,valentini2005dynamical}, $\rho_t(\vb x) \!=\! |\psi_t(\vb x)|^2$, and accomplishing normalization, the  joint arrival distribution could be represented by the absolute value of the current density as

\begin{equation}
   \mathbb{P}_{\text{QF}}(\vb x,t| \vb x\!\in\!\mathbb{S}) = \frac{ |\vb J(\vb x,t)\!\cdot\!\vb n|}{\int dt \int_{\mathbb{S}} dS  |\vb J(\vb x,t)\!\cdot\!\vb n|}. 
       \label{QFjoint}
\end{equation}
Now, by integrating \eqref{QFjoint} over all $\vb x\!\in\!\mathbb S$, we arrive at the three-dimensional arrival time distribution \eqref{QF3D} for the screen surface $\mathbb{S}$. It should be noted that Eq. \eqref{QFjoint} is not necessarily followed for an ensemble of classical particles because a positive or negative current at a space-time point, $(\vb{x}, t)$, can in general have contributions from all the particles arriving to $\vb{x}$ at $t$ from any direction. Nonetheless, since the Bohmian velocity field is single-valued, the particle trajectories cannot intersect each other at any point of space-time and so only a single trajectory contributes to the current density $\vb J(\vb x,t)$ at the particular space-time point $(\vb x,t)$. Moreover, this fact implies that when $\vb v (\vb x,t)  \cdot \vb n\!>\!0$ we can say that the trajectory and consequently the particle has passed through the screen from the inside and vice versa for $\vb v (\vb x,t) \cdot \vb n\!<\!0$. Hence, one can define the joint probability distribution for the time of arrival to each side of $\mathbb{S}$ as
\begin{equation}
   \mathbb{P}^\pm_{\text{QF}}(\vb x,t|\vb x\!\in\!\mathbb{S}) = \frac{ \vb J^\pm(\vb x,t)\!\cdot\!\vb n}{\int dt \int_{\mathbb{S}} dS \,\,  \vb J^\pm(\vb x,t)\!\cdot\!\vb n},
       \label{Ppm}
\end{equation}
where $\vb J^\pm ({\vb x},t)=\pm \, \theta (\pm\vb J\!\cdot\!\vb n)\vb \, \vb J(\vb x,t)$. 
 In addition, note that there may be some trajectories which cross $\mathbb{S}$ more than once—and we have \textit{multi-crossing} trajectories (see the typical Bohmian trajectory in Fig. \ref{SchematicSetup}). The course of the above inference to Eq. \eqref{QFjoint} was in such a manner that multi-crossing trajectories could contribute several times (see Fig.\,\ref{DetectionSchemes}\,(a)). However, one could assume the detection surface as a barrier that does not allow the crossed particle to return inside (see Fig.\,\ref{DetectionSchemes}\,(c)). In this case, it is suggested to use the truncated current defined as 
 \begin{equation}
	 { \tilde{\vb J}}({\vb x},t)
	 \coloneqq
	 \syst{ & {\vb J}({\vb x},t) && \mbox{if $({\vb x},t)$ is a first exit through $\mathbb{S}$}  
	\\ &0&& \mbox{otherwise} }  
	 \label{QFtruncated}
\end{equation}
where $({\vb x},t)$ is a first exit event through the boundary surface $\mathbb{S}$, if the trajectory passing through $\vb x$ at time $t$ leaves inside $\mathbb{S}$ at this time, for the first time since $t \!=\! 0$ \cite{daumer1997,Gebhard2002,vona2013does}. The limiting condition in \eqref{QFtruncated}, imposes that the joint probability distribution based on it should be computed numerically using trajectories:
 \begin{equation}
   \tilde{\mathbb{P}}_{\text{QF}}(\vb x,t|\vb x\!\in\!\mathbb{S}) = \frac{ \tilde{\vb J}(\vb x,t)\!\cdot\!\vb n}{\int dt \int_{\mathbb{S}} dS \,\,  \tilde{\vb J}(\vb x,t)\!\cdot\!\vb n}. 
       \label{truncatedP}
\end{equation}
 Of course, the detection screen is not always a barrier-like surface (see Fig.\,\ref{DetectionSchemes}\,(b)), and one could assume that there is a point-like detector that lets the multi-crossing trajectories to contribute to the distribution and we can use \eqref{QFjoint} in such cases. 

\section{``Intrinsic" screen observable in two-slit experiment}\label{sec3}
In this section, we study the discussed proposals in the previous section for the double-slit experiment. We compare the results of these proposals in the cases of vertical and horizontal screens (see Fig. \ref{SchematicSetup}), and also in different detection schemes. The main motivation for the study of the horizontal screen is the non-classical particles' motions along the $y$-direction, in the Bohmian perspective; see a typical Bohmian trajectory in Fig. \ref{SchematicSetup}. This behavior is due to changing the sign of the probability current's component in the $y$-direction. This behavior does not occur for $x$-component of $\vb J$ and consequently for the Bohmian motion of a particle along the $x$-direction. 

As shown in Fig. \ref{SchematicSetup}, the setup contains two identical slits at $y\!=\!\pm s$, and screens are placed at $x\!=\!L_x$ and $y\!=\!L_y$ correspond to the vertical and horizontal screens, respectively. To avoid the mathematical complexity of Fresnel diffraction at the sharp-edge slits, it is supposed that the slits have soft edges that generate waves having identical Gaussian profiles in the $y$-direction. So, for each slit, we can take the wave function as an uncorrelated two-dimensional Gaussian wave packet, which in each dimension has the form
\begin{align}\label{GaussianWave}
    \psi^{(i)}_G &(x,t)=(2\pi s_t^2)^{\textrm{-} \frac{1}{4} }\,\text{exp}\left\{\frac{(x-x^{(i)}_0-u_x t)^2}{4 \sigma_0 s_t}\right\}\nonumber\\
   &\times \text{exp}\left\{\frac{i}{\hbar}m u_x (x-x_0^{(i)}-\frac{u_x t}{2})\right\} \ \ (i=1,2),
\end{align}
with $m$ the particle's mass, $\sigma_0$ the initial dispersion, $u_x$ the wave packet's velocity, $x_0^{(i)}$ the initial position of wave packet or in other words the location of $i$-th slit, and $s_t=\sigma_0 (1+i \hbar t/(2m \sigma_0^2))$. Therefore, when the particle passes through the slits, we have the total wave function as
\begin{equation}\label{totalWaveFunction}
    \psi (x,y,t)=\frac{1}{\sqrt{2}}[\psi_G^{(1)} (x,t)\psi_G^{(1)}(y,t)+\psi_G^{(2)} (x,t)\psi_G^{(2)}(y,t)],
\end{equation}
where superscripts (1) and (2) correspond to upper and lower slits, respectively. This form of Gaussian superposition state is commonly used in the literature \cite{Hall2014Quantum,Viale2003Analysis,nitta2008time,Paul2017Measuring,Mishra2019Decoherence} and is feasible to implement by quantum technologies because such a state could be produced and controlled readily \cite{Fang2010Generation,laurat2005entanglement}, even without using slits \cite{Shin2004Atom}. Moreover, one can generalize Eq. \eqref{totalWaveFunction} by adding a relative phase between wave packets which is discussed in appendixes \ref{appendixA} and \ref{appendixB}. In this paper, we have chosen the metastable helium atom, with mass $m = 6.64 \times 10^{-27}$ kg, as the interfering particle, and the parameters as $s=10\,\mu$m, $\sigma_x=0.04\,\mu$m, $\sigma_y=0.5\,\mu$m, $u_x=3$ m/s, and $u_y=0$ m/s. These values are feasible according to the performed experiments \cite{Barnea2018Matter}. 
Moreover, the meta-stable helium atom could be detected with high efficiency because of its large internal energy \cite{khakimov2016ghost,Vassen2012Cold}. 

\subsection{Vertical screen}\label{VerticalScreen}
\begin{figure}[b]
    \centering
    \includegraphics[width = 8.6cm]{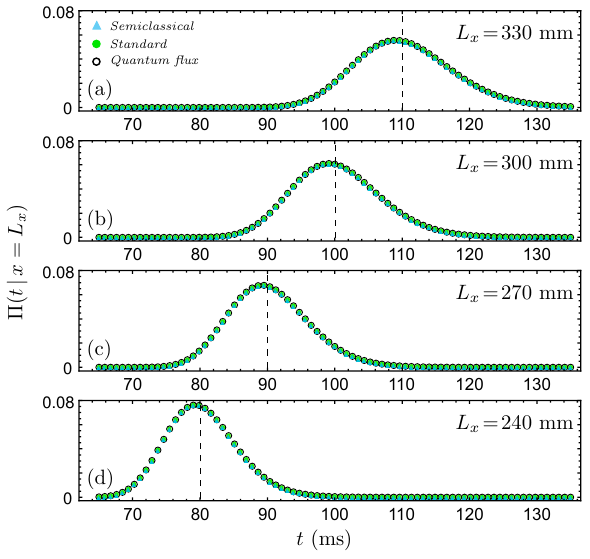}
    \caption{\textbf{Arrival time distributions at the vertical screen of the double-slit setup.} Panels (a), (b), (c), and (d) show the arrival distribution, $\Pi(t|x=L_x)$, at different screen distances $L_x=330,\,300,\,270,\,240\,$mm, respectively. The vertical dashed lines show the average arrival time obtained by the three methods, which have a perfect fit for all three methods. In all panels, the dark cyan triangle markers show the semiclassical approximation, the green-filled circle markers show the standard method and the black empty circle markers show the quantum flux approach.}
    \label{ATDV}
\end{figure}

The arrival time distribution for the vertical screen placed at different distances from the two-slit is shown in Fig.\,\ref{ATDV}. As one can see this distribution is the same for all methods, and their average arrival time is close to the corresponding quantity in classical uniform motion. To calculate the mean time of arrival to the screen, we use the arrival time distribution of each method presented in sec \ref{sec_Methods}, i.e., Eq. \eqref{SC2}, \eqref{3Dkij} and \eqref{QF3D}, and we have
\begin{equation}
    \bar t_{\mathbb S} =\int_0^{\infty}\!\!\! dt \,\, \Pi(t|\vb x\!\in\!\mathbb{S})\, t,
\end{equation}

 \begin{figure}[t]
        \centering
    \includegraphics[width = 8.6cm]{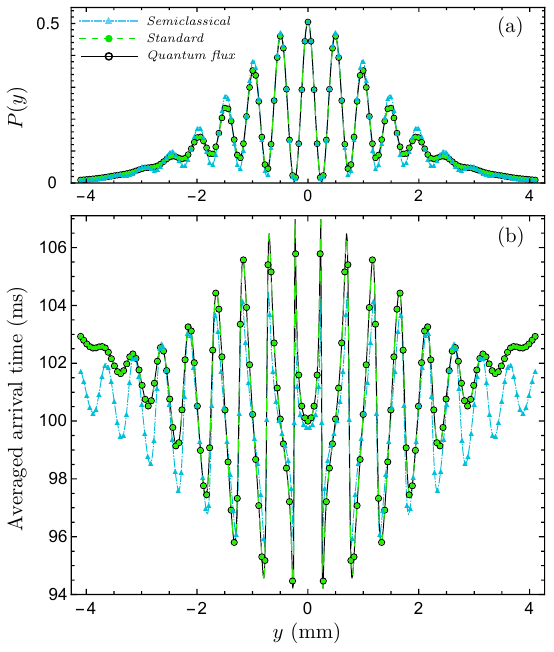}
    \caption{\textbf{Spatiotemporal arrival statistics for the double-slit experiment with a vertical screen.} Panel (a) represents the cumulative arrival position probability density. Panel (b) shows the average time of arrival at each point of the screen. In all panels, the dark cyan dot-dashed lines (with triangle markers) show the semiclassical approximation, the green dashed lines (with green-filled circle markers) show the standard method, and the black solid lines (with black empty circle markers) show the quantum flux approach.}
    \label{IPV-AATV}
\end{figure}

as the mean arrival time at the surface $\mathbb{S}$. Furthermore, we can compute the average arrival time to each point on the screen using the joint probability distribution as
\begin{equation}\label{averageTOA}
    \bar t _{\vb x} = \frac{\int_0^{\infty}dt\,\, \PP(\vb x,t|\vb x\!\in\!\mathbb{S})\,t }{\int_0^{\infty}dt\,\, \PP(\vb x,t|\vb x\!\in\!\mathbb{S}) }.
\end{equation}
This observable is depicted in Fig.\,\ref{IPV-AATV}-b for a vertical screen placed at $L_x\!=\!300$ mm. Apparently, the results of the \textit{standard} and \textit{quantum flux} methods are the same and similar to one that resulted in \cite{nitta2008time} by Nelson's mechanics. Nevertheless, they are different from the \textit{semiclassical} approximation. However, when the interference pattern is calculated by either method, we see that their predicted cumulative position distributions do not differ much from the others (Fig.\,\ref{IPV-AATV}-a). This observable can be calculated by using the joint distribution as
\begin{equation}\label{tempdis}
   P(\vb x | \vb x\!\in\!\mathbb{S}) \!=\! \int_0^{\infty}dt \,\, \PP(\vb x,t|\vb x\!\in\!\mathbb{S}).
\end{equation}
As mentioned, it should be noted that, $|\psi_t(\vb x)|^2$ is just the \textit{conditional} position probability density at the specific time $t$, not the position-time \textit{joint} probability density and so  the accumulated interference pattern, $P(\vb x | \vb x\!\in\!\mathbb{S})$, is not given by $\int dt |\psi_t(\vb x)|^2$ \cite{das2022double}. 

\begin{figure}[b!]
    \centering
    \includegraphics[width = 8.6cm]{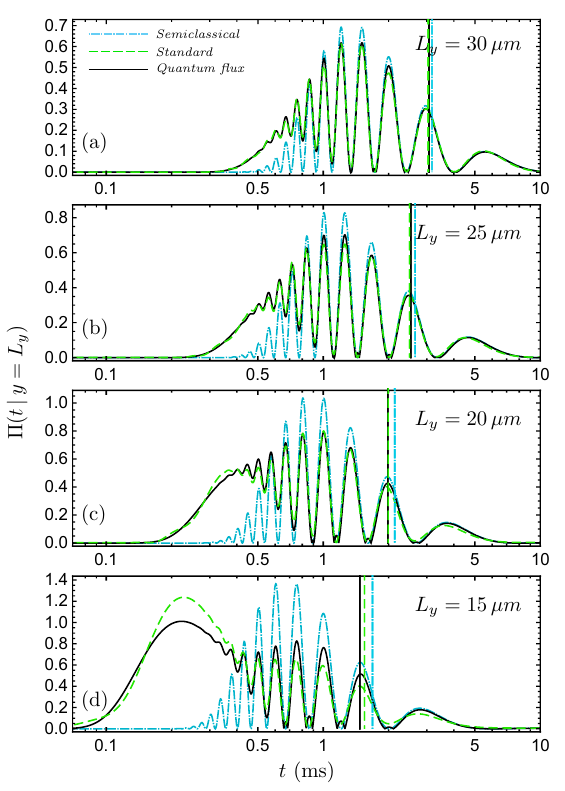}
   \caption{\textbf{Arrival time distributions at the horizontal screen of the double-slit setup.} Panels (a), (b), (c), and (d) show the arrival distribution, $\Pi(t|y=L_y)$, at different screen distances $L_y=30,\,25,\,20,\,15\,\mu m$, respectively. The vertical lines show the average arrival time. In all panels, the dark cyan dot-dashed lines show the semiclassical approximation, the green dashed lines show the standard method, and the black solid lines show the quantum flux approach.}
    \label{ATDH_1D_AB}
\end{figure}

\subsection{Horizontal screen} \label{HorizontalScreen}
In this section, we are going to compare the mentioned proposals in the double-slit setup with a horizontal detection screen (see Fig.\,\ref{SchematicSetup}).  
In this regard, in  Fig.\,\ref{ATDH_1D}, the arrival time distributions at the screen are plotted for some horizontal screens which are located at  $L_y\!=\!15$, $\,20$, $\,25$, and $\,30\, \mu$m. In this figure, solid-black, dashed-green, and dash-dotted-blue curves  represent the distributions $\Pi_{STD}$, $\Pi_{QF}$ and $\Pi_{SC}$ respectively. Also, the vertical lines show the average time of arrival to the screen, $\bar t_{\mathbb{S}}$, associated with these arrival time distributions.
From this figure, one can see that, although the averages almost coincide, the distributions are distinct. Moreover, as expected,  when the screen's distance from the center of the two slits $L_y$ decreases, the difference between distributions increases.  Most of these differences occur in the early times, which are associated with the particles that arrive at the $\mathbb{S}$ in the near field. Furthermore, we observe that the $\Pi_{SC}$ behaves quite differently from $\Pi_{QF}$ and $\Pi_{STD}$. The distributions $\Pi_{QF}$ and $\Pi_{STD}$  are more or less in agreement, however, for the screen that is located at $L_y\!=\!15\,\mu$m, a significant difference between the standard and quantum flux distributions occurs around $t\!\approx\!0.2$ ms. 

\begin{figure}[h!]
    \centering
    \includegraphics[width = 8.6cm]{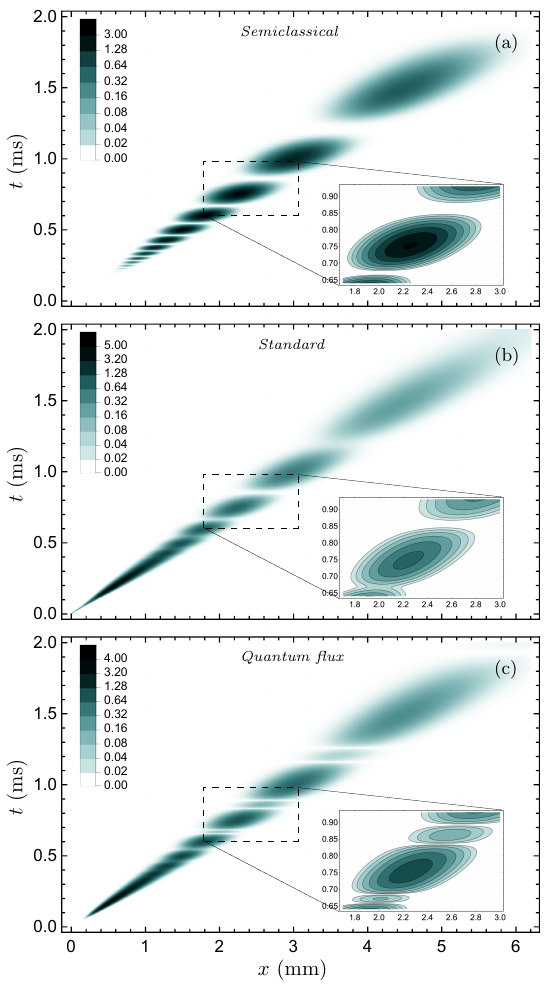}
    \caption{\textbf{Joint spatiotemporal probability distributions on the horizontal screen of the double-slit setup.} Panels (a), (b), and (c) represent intrinsic distributions predicted by the semiclassical approximation, standard method, and quantum flux approach, respectively. Insets: Magnified contour plots of the joint distributions.}
    \label{ScreenObservables}
\end{figure}

To have a more comprehensive insight, we can look at the joint spatiotemporal arrival distributions in Fig.\,\ref{ScreenObservables}. In this figure, joint distributions, $\mathbb{P}_{\text{SC}}$, $\mathbb{P}_{\text{STD}}$ and $\mathbb{P}_{\text{QF}}$ are plotted in three panels, for the horizontal screen surface located at $L_y\!=\!15\,\mu$m. These density plots clearly visualize differences between the mentioned arrival distribution proposals. In these plots, we can see separated fringes with different shapes, which this fact imply that the particles arrive at the screen in some detached space-time regions. In the insets, one can see that the shapes of these regions are different for each proposal.  In the  joint density of the  semiclassical approximation (Fig.\,\ref{ScreenObservables}-a), fringes are well-separated, while  the standard distribution (Fig.\,\ref{ScreenObservables}-b) exhibits more continuity in its fringes. In addition, in the pattern of the quantum flux proposal (Fig.\,\ref{ScreenObservables}-c) there are grooves between every two fringes which is due to changing the sign of $\vb{J}(\vb x,t) \cdot\vb{n}$ in \eqref{QFjoint}. In all panels of Fig.\,\ref{ScreenObservables}, the duration of ``temporal no-arrival windows" between every two typical fringes variate in the range between $0.01$ and $0.2$ ms which has a spatial extension of about $0.3$ to $2$ mm. These space-time scales are utterly amenable empirically by current technologies \cite{Barnea2018Matter,kurtsiefer19962A}, which could be used to test these results. 

\begin{figure*}
    \centering
    \includegraphics[width = 18cm]{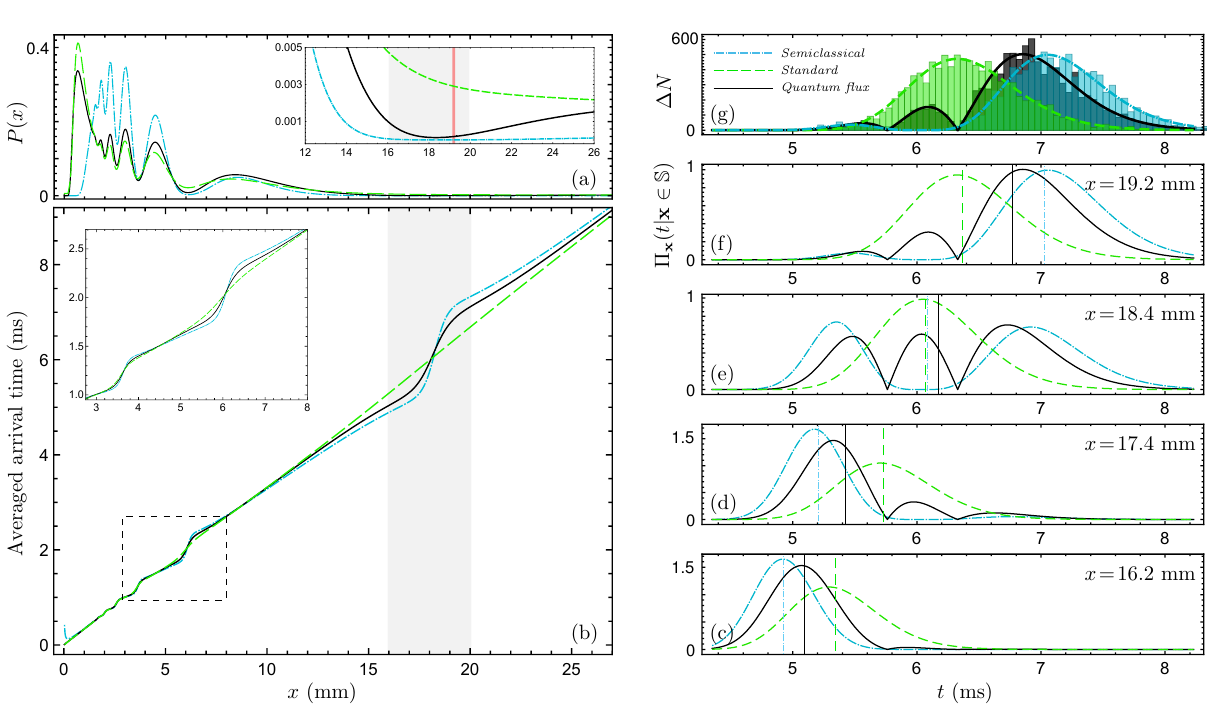}
    \caption{\textbf{Spatiotemporal arrival statistics for the double-slit experiment with a horizontal screen.} Panel (a) represents the cumulative arrival position probability density. The inset of this panel is a zoom-in of the main plot close to the horizontal axis. Panel (b) shows the average time of arrival at each point of the screen. The inset of this panel is a zoom-in of the area marked with the dashed square in the main plot. Panels (c), (d), (e), and (f) show the local arrival time probability densities, $\Pi_{\vb x} (t|\vb x \!\in \! \mathbb{S})$, at the points $x\!=\! 16.2,\,17.4,\,18.4,\, 19.2 $ mm on the screen $\mathbb{S}$ placed at $y\!=\!15 \, \mu$m, respectively, which are chosen from the gray region in panel (b). The vertical lines in these panels represent the average arrival times. Panel (g) contains the Histograms associated with the probability densities of the panel (f), which are generated by $10^4$ numerical random sampling. In all panels, the dark cyan dot-dashed lines show the semiclassical approximation, the green dashed lines show the standard method, and the black solid lines show the quantum flux approach.}
    \label{AATH}
  \end{figure*}

The average time of arrival to each point of the screen and cumulative position interference pattern could be calculated as in the vertical screen case by Eqs.\,\eqref{averageTOA} and \eqref{tempdis}. In Fig.\,\ref{AATH}(a)-(b),  these two quantities are shown for the horizontal screen which is placed at $y\!=\!15\,\mu$m. In contrast to the vertical screen, the cumulative position distribution of the semiclassical approximation is entirely separate from the two other proposals. The cumulative position distribution resulting from standard and quantum flux approaches have obvious differences from each other, as well. 
As one can see in Fig.\,\ref{AATH}(b), the average arrival times are the same for all three methods at first and begin to deviate from each other at $x\!\approx\!5$ mm; then again, these curves converge to each other at $x\!\approx\!25$ mm, approximately. The maximum deviation between the standard and quantum flux average arrival time occurs at $x\!\approx\!19$ mm, which is quite in the far-field regime---the width of the initial wave function is $\sim O(10^{-3})$mm which is smaller than $19$ mm. Therefore one can suggest the average arrival time in the gray region of Fig.\,\ref{AATH}(b) as a practical target for comparing these approaches experimentally.
To this end, we study arrival time distributions at some points of this region as \textit{local arrival distributions}. The arrival time distribution conditioned at a specific point $\vb x$ on the screen can be obtained as follow
\begin{equation}\label{eqATDH}
    \Pi_{\vb x} (t|\vb x \!\in \! \mathbb{S}) = \frac{ \PP(\vb x,t|\vb x\!\in\!\mathbb{S})}{\int_0^{\infty}dt \,\, \PP(\vb x,t|\vb x\!\in\!\mathbb{S}) }.
\end{equation}
Using the associated joint distribution of each proposal, we have plotted Fig.\,\ref{AATH}(c)-(f) that show $\Pi_{\vb x} (t|\vb x \!\in \! \mathbb{S})$ at the positions $x\!=\! 16.2,\,17.4,\,18.4,\, 19.2 $ mm, on the screen placed at  $L_y\!=\!15 \,\mu$m. The broken black curves in Fig.\,\ref{AATH}\,(c)-(f), resulting from the quantum flux proposal, against the smooth curves of the other two methods could be understood as the result of the changing the signature of the $y$-component of the probability current: Note that, quantum flux distribution is given by the absolute value of  the probability current.  The origin of distinctions between the local average arrival times is more perceptible from these local arrival distributions. In principle, these distributions could be probed using fast and high-resolution single-atom detectors \cite{kurtsiefer19962A,Vassen2012Cold}.
In particular, the delay-line detector that is recently developed by Keller  et al. \cite{Keller2014Bose} seems suitable for our purpose: It has the capability to resolve single-atom detection events temporally with $220$\,ps and spatially with $177\mu$m at rates of several $10^6$ events per second. 

We estimate by a numerical investigation that these local arrival distributions could be well reconstructed from about $10^4$ number of detection events.
As an example, in Fig.\,\ref{AATH}, the histograms associated with the probability densities of the panel (f) are plotted in panel (g), using $10^4$ numerical random sampling. It is easy to estimate that the recording of $10^4$ particle detection events can determine the local average arrival time with a statistical error of about $10^{-2}$ms, while the differences between local average arrival times of various proposals are almost bigger than $10^{-1}$ms. Using cumulative position distribution, Fig.\,\ref{AATH}(a), one can estimate that, if the total number of particles that arrived at the screen is about $10^8$, we have about $10^4$ particles around $x=19.2$\,mm, in the spacial interval $(19.1, 19.3)$. Using recent progress in laser cooling and magneto-optical trapping \cite{Vassen2012Cold}, the preparation of a coherent ensemble of metastable helium atoms with this number of particles is quite achievable \cite{Keller2014Bose}.

\begin{figure*}
    \centering
    \includegraphics[width = 18cm]{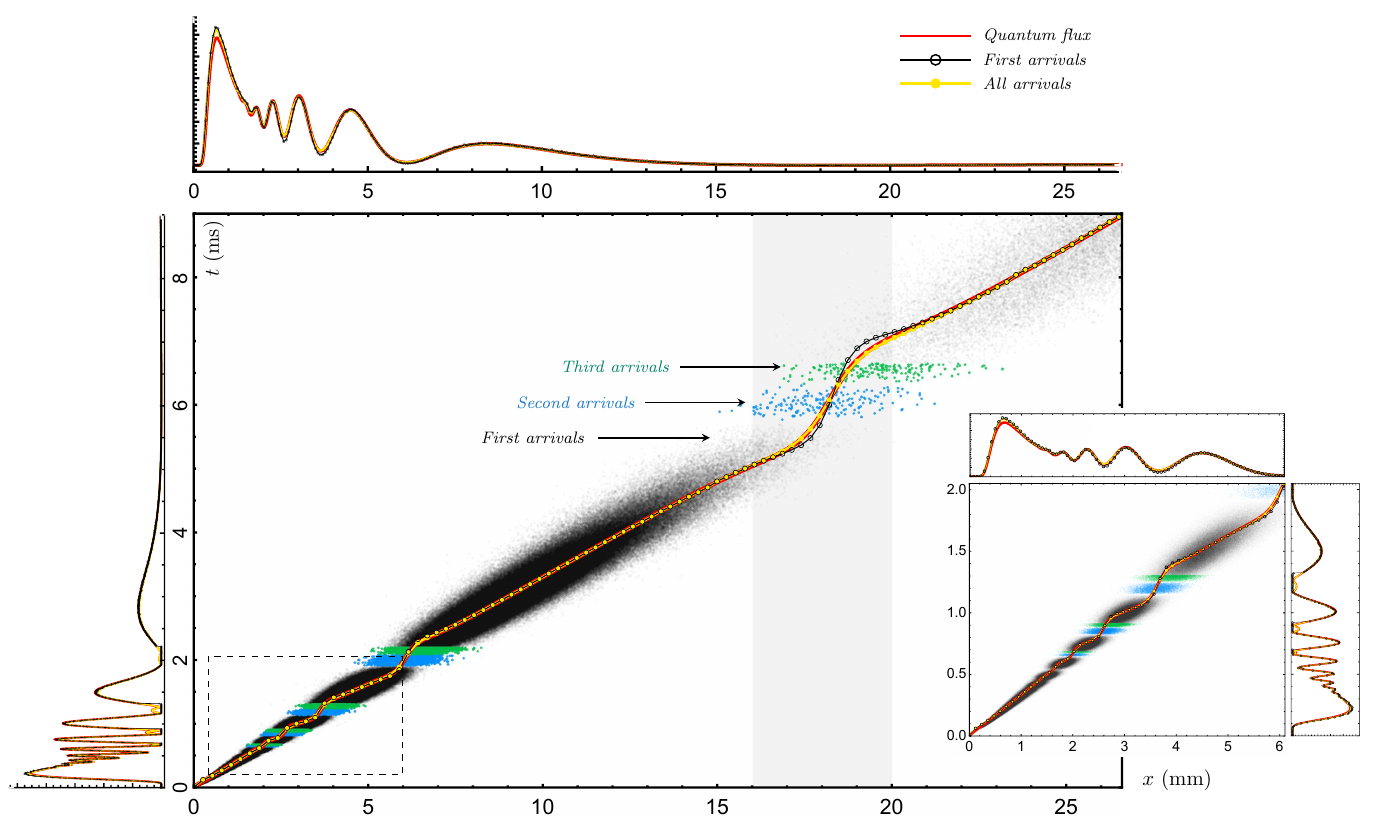}
   \caption{\textbf{Spatiotemporal Bohmian arrival statistics for the double-slit experiment with a horizontal screen.} The interior curves in the central figure are the averaged times of arrival obtained by different detection schemes (see Fig.\,\ref{DetectionSchemes}) using $10^8$ simulated Bohmian trajectories. The left and top plots are marginal arrival time distributions and marginal arrival position distributions, respectively. The scatter plot is generated using $2\times 10^6$ Bohmian trajectories, and the black, blue, and green points of the scatter plot represent the first, second, and third arrivals of Bohmian particles to the screen, respectively. The inset is a zoom-in of the dashed rectangle. The red solid lines represent the quantum flux approach, the black solid lines with empty circles markers show the first-arrivals scheme, and the yellow solid lines with filled circles show the all-arrivals scheme.}    \label{AATH_IPH_Numeric}
\end{figure*}

 One might be inclined to think that the difference between the quantum flux and standard average arrival times is just due to  changing the signature of $\vb{J}(\vb x,t)\cdot \vb{n}$, but in the following, we show that even without the contribution of the negative part of $\vb{J}(\vb x,t)\cdot \vb{n}$, these proposals are significantly distinguishable: see Fig.\,\ref{AATH_IPH_Numeric}.

\subsection{Detection schemes} \label{Detectionschemes}

As we mentioned in section \ref{subsec_QF}, according to the Bohmian deterministic point of view, there are several possible schemes to detect arrived particles, especially for the horizontal screen surface which we have recursive motions (see Fig.\,\ref{SchematicSetup} and \ref{DetectionSchemes}). One can assume that the horizontal screen is swept with a point-like detector that surveys all arrived particles at the surface $\mathbb S$, which we call \textit{spot-detection} scheme. In this scheme, one option is to use a unilateral detector to detect arrived particles at the top or bottom of $\mathbb S$. In this case, the positive and negative parts of the quantum probability current have respectively corresponded to particles that arrive at the top or bottom of $\mathbb S$ (as shown in Fig.\,\ref{DetectionSchemes}\,(a)), and we must use Eq.\,\eqref{Ppm} to calculate the screen observables. Additionally, we can choose a bilateral detector (or two unilateral detectors) that probe all particles that arrive from both sides of $\mathbb S$, along the time with several repeats of the experiment (as shown in Fig.\,\ref{DetectionSchemes}\,(b)). In these circumstances (i.e. spot-detection scheme), there is no barrier in front of the particles before they reach the point of detection and we can use Eq. \eqref{QFjoint} to obtain the screen observables as in the two previous subsections.

As we have already shown in section \ref{subsec_QF}, whether the particles arrive from the top or bottom of $\mathbb S$, the absolute value of the quantum probability current yield the trajectories' density and consequently give the joint distribution of the total arrival at each point of $\mathbb S$. This fact is the case for the standard method, as well, however, there is a subtle difference between the two proposals in the spot-detection scheme. When we talk about the spot-detection in the Bohmian approach, it would be considered the possibility of multi-crossing and the distribution includes \textit{all-arrivals} at $\mathbb S$. Although, in the standard method there is an interpretation for $\psi^+_\mathbb{S}(\vb x,t)$ and $\psi^-_\mathbb{S}(\vb x,t)$ in Eq. \eqref{psiSpm}, which relates them to the particles arrive at $\mathbb S$ in a direction which is the same or opposite with the direction of outward normal of the screen $\vb n$, respectively \cite{Kijowski1974,art:Grot-Rovelli-Tate}. Nevertheless, it should be noted that, in contrast to the Bohmian interpretation, since there is no defined particle trajectory in the standard interpretation, it is meaningless to ask whether it only counts the \textit{first arrivals} to each side of the screen or includes recursive movements of particles. In standard quantum mechanics, there is only one arrival because once measurement has been made the state of the particle is not causally connected to the initial state \cite{flores2019quantum}.

\begin{figure}[t]
  \centering
  \includegraphics[width = 8.6cm]{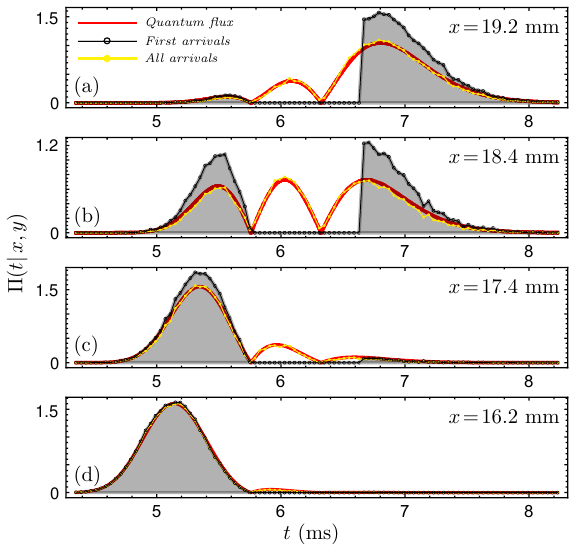}
  \caption{\textbf{Local arrival time distribution at some points of the horizontal screen in the double-slit setup.} Panels (a), (b), (c), and (d) show the local arrival time probability densities, $\Pi_{\vb x} (t|\vb x \!\in \! \mathbb{S})$, at the points $x\!=\! 16.2,\,17.4,\,18.4,\, 19.2 $ mm on the screen $\mathbb{S}$ placed at $y\!=\!15 \, \mu$m, respectively, which are in the gray region of Fig (\ref{AATH_IPH_Numeric}). The width of sampling in each point is about $\delta x =0.25$ mm, and $10^8$ Bohmian trajectories are simulated to obtain these distributions.}
  \label{ATDH_Numeric}
\end{figure}

Alternatively, along with the spot-detection scheme, it could be assumed that there is a continuous flat barrier in front of the particle's paths as the \textit{detection surface} or \textit{screen} surface that does not allow particles to cross this surface.
Depending on the screen's length and position, there are several possibilities for the detection process. In each case, a specific number of particle paths contribute to the distribution of arrival time. In the simplest case, the screen blocks all the trajectories that reach the horizontal surface $\mathbb S$, and we only detect the first-arrivals. In such a setup, we can no longer use the quantum flux method to represent Bohmian trajectories' first encounter with the surface; hence, the screen observables must be obtained by numerical analysis, due to the definition of truncated current as in Eq. \eqref{QFtruncated} and its corresponding joint distribution, $\tilde\PP_{\text{QF}}(\vb x,t|\vb x\!\in\!\mathbb{S})$, defined in Eq. \eqref{truncatedP}. By computing the Bohmian trajectories, we can find positions and times of the first-arrivals to the screen, and consequently calculate the arrival time distribution which mathematically could be defined as
\begin{equation}
  \tilde\Pi_{\text {QF}}(t | \vb x \! \in \! \mathbb{S}) = \int_{\mathbb{S}}\tilde\PP_{\text{QF}}(\vb x,t|\vb x\!\in\!\mathbb{S}) dS.
\end{equation}
Also, other observable quantities such as the cumulative spatial distribution and averaged arrival time over the detection surface could be defined and calculated numerically in a similar way—by substituting $\tilde\PP_{\text{QF}}(\vb x,t|\vb x\!\in\!\mathbb{S})$ in Eqs. \eqref{averageTOA} and \eqref{tempdis}. Furthermore, we can complete the computations to find the second and third encounters to the surface (regardless of the barrier). 

In Fig.\,\ref{AATH_IPH_Numeric}, we show our numerical results of Bohmian trajectories simulation. The background scatter plot is the position and time of arrivals of $2\times10^6$ trajectories. In this plot, the second and third arrivals are shown in blue and green, respectively. Here, it is more clear why we interpret the grooves of the quantum flux density plot (Fig.\,\ref{ScreenObservables}\,(c)) as a result of the multi-crossing of Bohmian trajectories. The three middle graphs are the average time of the \textit{first} and all-arrivals, which are simulation results of $10^8$ trajectories, and are compared by the quantum flux method. As expected, the average time of all-arrivals fits on the quantum flux curve. However, the average time of first-arrivals deviates from all-arrivals in the area discussed in the previous section (between $x=16.2$ mm and $x=19.2$ mm). 
  
To scrutinize the deviation zone of Fig.\,\ref{AATH_IPH_Numeric} (the gray region), Fig.\,\ref{ATDH_Numeric} is drawn to show the arrival time distributions of screen positions $x\!=\! 16.2,\,17.4,\,18.4,\, 19.2 $ mm. As one can see, at the first recursive points of quantum flux distribution, the first-arrival distributions raise down to zero. This implies that in the presence of a barrier-like screen, there would be a big temporal gap in the local arrival distribution at these points. These gaps could be investigated as a result of the non-intersection property of Bohmian trajectories that cause a unilateral motion of particles along the direction of the probability current field. 

\section{Screen back-effect}\label{screen_back_effect}
In principle, the presence of the detector could modify the wave function evolution, before the particle detection, which is called detector back-effect.
To have a more thorough investigation of detection statistics, we should consider this effect. However, due to the measurement problem and the quantum Zeno effect \cite{Allcock1969,misra1977zeno,porras2014quantum}, a complete investigation of the detector effects is problematic at the fundamental level, and it is less obvious how to model an ideal detector 
\footnote{ See some recent interesting papers deal with these problems especially in connection with arrival time problem \cite{juric2022passive,Dubey2021Quantum} }. 
 Nonetheless, some phenomenological non-equivalent models are proposed, such as the generalized Feynman path integral approach in the presence of absorbing boundary \cite{Marchewka1998Feynman,Marchewka2000Path,Marchewka2001Survival,Marchewka2002}, Schrödinger equation with a complex potential \cite{tumulka2022absorbing}, Schrödinger equation with absorbing (or complex Robin) boundary condition \cite{werner1987arrival,Tumulka2022Distribution,Tumulka2022Detection,Dubey2021Quantum,tumulka2022absorbing}, and so on \cite{juric2022passive}. The results of these approaches are not the same, and a detailed study of the differences is an interesting topic. In this section, we provide a brief review of the absorbing boundary rule (ABR) and path-Integral with absorbing boundary (PAB) models, then we compare them in the double-slit setup with the horizontal screen. 
 
  \subsection{Absorbing Boundary Rule}
Among the above-mentioned phenomenological models, the absorbing boundary condition approach has the most compatibility with Bohmian mechanics \cite{Tumulka2022Detection}. The application of absorbing boundary condition (ABC) in arrival time problem was first proposed by Werner \cite{werner1987arrival}, and recently it is re-derived and generalized by Tumulka and others using various methods \cite{Tumulka2022Distribution,Tumulka2022Detection,Dubey2021Quantum,tumulka2022absorbing}. Especially, it is recently shown that in a suitable (non-obvious) limit, the imaginary potential approach yields the distribution of
detection time and position in agreement with the absorbing boundary rule \cite{tumulka2022absorbing}. According to this rule, the particle wave function $\psi$ evolves according to the free Schr\"{o}dinger equation, while the presence of a detection screen is modeled by imposing the following boundary conditions on the detection screen, $\vb{x}\in \mathbb{S}$, 
\begin{equation}\label{ABC_eq}
\vb n\cdot\nabla\psi=i\kappa\psi,
\end{equation}
 where $\kappa \!>\!0$ is a constant characterizing the type of detector, in which $\hbar \kappa/m$ represents the momentum that the detector is most sensitive to. This boundary condition ensures that waves with wave number $\kappa$ are completely absorbed while waves with other wave numbers are partly absorbed and partly reflected \cite{Tumulka2022Distribution,Fevens1999Absorbing}. In the absorbing boundary rule, the joint spatiotemporal distribution of the detection event is given by quantum flux. Considering \eqref{ABC_eq}, this distribution reads
\begin{equation}
   \mathbb{P}_\text{ABR}(\vb x , t|\vb x\!\in\!\mathbb{S}) = \frac{|\psi_\text{ABC}|^2}{\int dt \int_{\mathbb{S}} dS  |\psi_\text{ABC}|^2},
    \label{ABC_joint}
\end{equation} 
 where $\psi_\text{ABC}$ represent the solution of the free Schr\"{o}dinger equation satisfying the aforementioned absorbing boundary condition.  This distribution can be understood in terms of Bohmian trajectories. The Bohmian particle equation of motion, 
 $\dot{\vb{X}}=(\hbar/m) \text{Im}\left[\nabla \psi_\text{ABC}/\psi_\text{ABC}\right]$,
together with the boundary condition \eqref{ABC_eq}, imply that trajectories can cross the boundary $\mathbb{S}$ only outwards and so there are  no multi-crossing trajectories. If it is assumed that the detector clicks when and where the Bohmian particle reaches $\mathbb{S}$, the probability distribution of detection events is given by \eqref{ABC_joint}, because the initial distribution of the Bohmian particle is $|\psi_\text{ABC}(x,0)|^2$ \cite{Tumulka2022Distribution}.
\begin{figure}[b]
  \centering
  \includegraphics[width = 8.6cm]{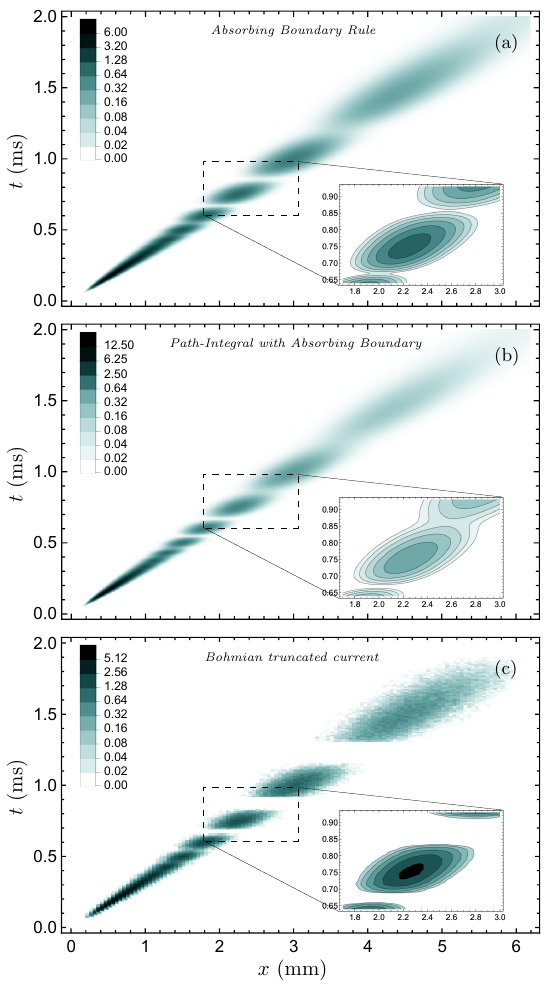}
  \caption{\textbf{Joint spatiotemporal probability distributions on the horizontal screen of the double-slit setup.} Panels (a) and (b) are the joint arrival distributions predicted by methods that take the screen back-effect into account, i.e., absorbing boundary role and path integral with absorbing boundary, respectively. Panel (e) is the joint arrival distribution obtained by Bohmian truncated current that is calculated numerically by simulating $10^7$ Bohmian trajectories. Insets: Magnified contour plots of the joint distributions.}
  \label{ScreenEffects}
\end{figure}

\subsection{Path-Integral with Absorbing Boundary}
In several papers \cite{Marchewka1998Feynman,Marchewka2000Path,Marchewka2001Survival,Marchewka2002}, Marchewka and Schuss develop an interesting method to calculate the detection effect of absorbing surface using the Feynman path integral method. They postulate a \textit{separation principle} for the wave function in which we could consider the (bounded wave function) as a sum of two parts, $\psi(\vb x, t)=\psi_1(\vb x, t)+\psi_2(\vb x, t)$, such that $\psi_1(\vb x, t)$ corresponds to the survival part of the wave which is orthogonal to $\psi_2(\vb x, t)$ at a time $t$ and evolve independently \cite{Marchewka2000Path}. So, we can obtain the probability of survival of the particle, denoted $S(t)$, which is the probability of the particle not being absorbed by the time $t$, as $\int_\mathbb {D} d^3\vb x|\psi_1(\vb x,t)|^2 $, where the integral  is over the domain $\mathbb D$, outside the absorbing region. By discretizing the path integral in a time interval $[0,t]$ and eliminating the trajectories that, in each time interval $[t',t'+\Delta t']$ for all $t'<t$, are reached to the absorbing surface $\mathbb S$, the survival and consequently absorbing probability would be obtained. Based on this analysis, we could define a unidirectional probability current into the surface as $\frac{d}{dt}[1-S(t)]$, which yields a normal component of the multidimensional probability current density at any point on $\mathbb S$ as
\begin {eqnarray}
	\vb J(\vb x,t)\!\cdot\! \vb n&\!\!=\!\!&\frac{\lambda \hbar}{m \pi} \,|\vb n \!\cdot\! \grad \psi (\vb x,t)|^2  \nonumber \\
	&&\!\times\,\, \text {exp} \left\{-\frac{\lambda \hbar}{m \pi} \int_0^{t} \!\! dt'\! \oint_{\mathbb S}dS |\vb n \!\cdot\! \grad \psi (\vb x',t')|^2\right\},\nonumber\\
\end{eqnarray}
where $dS\!=\!\vb n\cdot d\vb S$ is the magnitude of the surface element $d\vb S$, $\vb n$ is the unit outer normal to the absorbing surface $\mathbb S$, and $\lambda$ is a proportionality factor with the dimension of length \cite{Marchewka1998Feynman,muga2000arrival}. Also, $ \psi (\vb x,t)$ is the solution of Schrödinger equation bounded and normalized in the domain $\mathbb D$. Moreover, the normal component $\vb J(\vb x,t)\!\cdot\! \vb n$ is supposed to be the probability density for observing the particle at the point $\vb x$ on the screen at time $t$ \cite{Marchewka2001Survival,Marchewka2002}.

\begin{figure}[b] 
  \centering
  \includegraphics[width = 8.81cm]{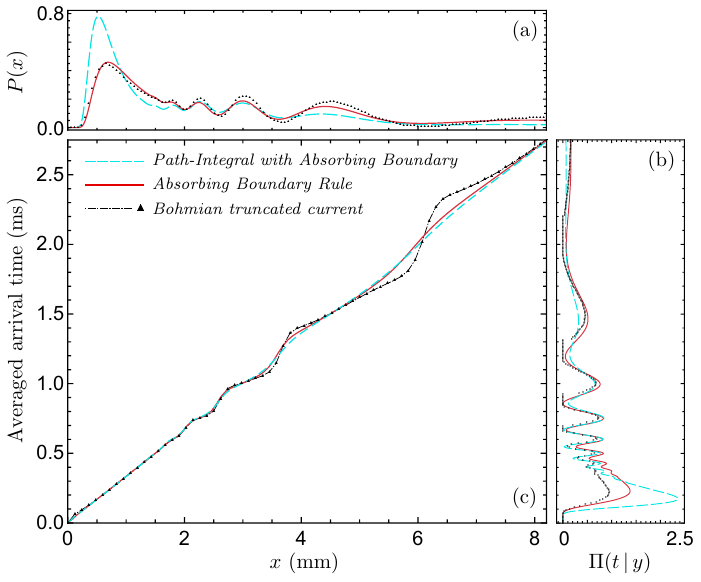}
  \caption{\textbf{Spatiotemporal arrival statistics for the double-slit experiment with absorbing horizontal screen.} Panel (a) represents cumulative interference patterns, panel (b) shows arrival time distributions on the screen, and plots in panel (c) are the averaged time of arrival at each point of the screen. In all panels, the red solid lines represent the absorbing boundary rule, the cyan dashed lines correspond to the path-integral with absorbing boundary method, and black triangle markers belong to the plots obtained by Bohmian truncated current method.}
  \label{AATH_IPH_ATD}
\end{figure}
\begin{figure}[t]
  \centering
  \includegraphics[width = 8.6cm]{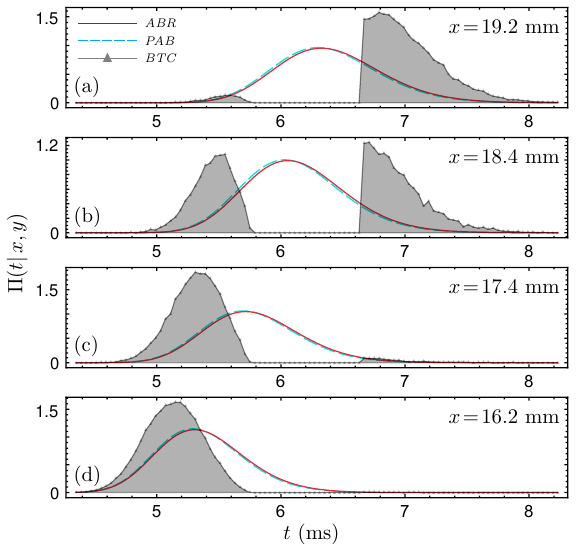}
  \caption{\textbf{Local arrival time distribution at some points of the horizontal screen in the double-slit setup.}  Panels (a), (b), (c), and (d) show the local arrival time probability densities, $\Pi_{\vb x} (t|\vb x \!\in \! \mathbb{S})$, at the points $x\!=\! 16.2,\,17.4,\,18.4,\, 19.2 $ mm on the screen $\mathbb{S}$ placed at $y\!=\!15 \, \mu$m, respectively. The black solid lines with triangle markers show the first-arrival scheme which obtained from Bohmian truncated current (BTC). The width of numeric sampling of this scheme in each point is about $\delta x =0.25$ mm, and $10^8$ Bohmian trajectories are simulated to obtain these distributions. The red solid lines represent the absorbing boundary rule (ABR), and the cyan dashed lines show the local arrival distributions obtained by path-integral with absorbing boundary (PAB).}  \label{ATDH_ScreenEffects}
\end{figure}

\subsection{Screen back-effect in two-slit experiment}
In order to complete the investigations carried out in section \ref{sec3}, we are going to study the screen back-effect in the double-slit experiment with a horizontal screen. In this regard, we compare the arrival distributions which are resulted from the absorbing boundary rule (ABR), path-Integral with absorbing boundary (PAB), and Bohmian truncated current (BTC).

We continue with the same initial conditions as in section \ref{sec3}, and choose $\kappa\!=\!1\,\mu\text{m}^{-1}$ for ABR. This value of $\kappa$ leads to the maximum absorption probability---which is almost $0.4$---for the chosen initial wave function. In addition, for a more meaningful comparison, we consider $\lambda\!=\!1\,\mu$m in the PAB method, which leads to the same absorption probability as ABR. The resulting joint arrival time-position distributions of the three methods are depicted in Fig.\,\ref{ScreenEffects}. As one can see, the distributions of the ABR and PAB methods—i.e., panels (a) and (b) in Fig.\,\ref{ScreenEffects}—have more compatibility with each other than the result of the BTC method. However, there are differences between them which are more obvious in the zoomed areas. The joint density of the ABR is more uniformly distributed than of the PAB method. The empty areas between the fringes of the panel (c) of Fig.\,\ref{ScreenEffects} are due to the elimination of the recursive trajectories—or in other words, are due to the elimination of second and third arrivals in Fig.\,\ref{AATH_IPH_Numeric}.

For a more detailed comparison, in Fig.\,\ref{AATH_IPH_ATD} the spatial and temporal marginal distributions are shown. In addition, the associated local average arrival times are compared in the central panel of this figure. The PAB method leads to significant discrepancies in marginal distributions; The maximum difference is about $40\%$ that occurs around $x\!\approx\!0.8$ mm, which seems testable clearly.
In contrast to the previous results on intrinsic distributions, in which the difference between average arrival times was significant, there is a good agreement in this observable for the ABR and PAB methods. However, there is a significant difference between the average arrival time in these two methods and BTC around $x\!=\!6$ mm. In Fig.\,\ref{ATDH_ScreenEffects}, the local arrival time distributions at some points on the screen are plotted, which show similar behavior.

\section{Summary and Discussion}\label{summary}
When and where does the wave function collapse? How one can model a detector in quantum theory? These are the questions that we investigated in this work. We tried to show that there is no agreed-upon answer for these questions, even for the double-slit experiment that \textit{has in it the heart of quantum mechanics} \cite{aharonov2017finally}. This is a practical encounter with the measurement problem \cite{galapon2009theory}. In this regard, we numerically investigated and compared the main proposed answers to these questions for a double-slit setup with a horizontal detection screen. It is shown that these proposals lead to experimentally distinguishable predictions, thanks to the current single-atom detection technology. 

In this work, we suggest the meta-stable helium atom as a proper coherent source of the matter wave, however, other sources may lead to some practical improvements. For example, using heavier condensate atoms can lead to more clear discrepancies. Recently, S. Roncallo and coworkers suggest an interesting experiment, using the $^{87}\text{Rb}$ Bose-Einstein condensate trapped in an accelerator ring \cite{pandey2019hypersonic}, to probe the various arrival time proposals \cite{roncallo2023does}. Moreover, pairs of entangled atoms, for example in a double-double slit setup, may lead to predictions that are more distinguishable \cite{kazemi2022detection,anastopoulos2017time,Rafsanjani2023Non}.

Finally, it is worth noting that, although the experiment with photons may have some practical advantages, there are more complications in its theoretical analysis. This is partially because of the relativistic localization-causality problem \cite{Hegerfeldt1985Violation,sebens2019electromagnetism,Kazemi2018Probability,Terno2014}. The theoretical investigation of the proposed experiment for photons would be an interesting extension of the present work, which has been left for future studies.

\begin{acknowledgments}
We sincerely thank  S. Goldstein, S. Das, M. Khorrami, L. Maccone, H. M. Wiseman, H. Nikolic, 
W. Struyve, E. Galapon, K. Sacha, and M. H. Barati for their kindly feedback and helpful comments.  
\end{acknowledgments}

\appendix

\section{  Effects of the relative phase}\label{appendixA}
One might propose that the difference between approaches could be increased by choosing a proper relative phase $\varphi$ in the initial wave function as follow:
\begin{equation}\label{WF}
    \psi (x,y,t)\!=\! \frac{1}{\sqrt{2}}[\psi_G^{(1)} (x,t)\psi_G^{(1)}(y,t)+\text {e}^{i\varphi} \psi_G^{(2)} (x,t)\psi_G^{(2)}(y,t)],\nonumber
\end{equation}
This relative phase could be produced and controled using the Aharonov-Bohm effect for charged particles \cite{Aharonov1959Significance,Batelaan2009TheAharonov}, the Aharonov-Casher effect \cite{Aharonov1984Topological} for neutral particles \cite{aitchison1989topological,Carnal1991Young}, and other methods \cite{Perreault2005Observation}.
\begin{figure}[h!]
    \centering
    \includegraphics[width = 8.6cm]{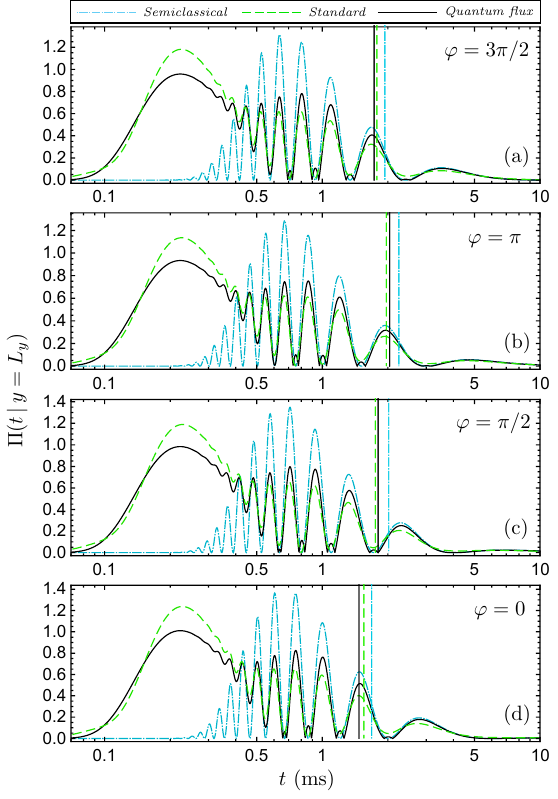}
    \caption{\textbf{Arrival time distributions at the horizontal screen of the double-slit setup.} Panels (a), (b), (c), and (d) show the arrival distribution, $\Pi(t|y=L_y)$, at screen distances $L_y\!=\!15\,\mu$m for four different relative phases $\phi=1.5\,\pi,\,\pi,\,0.5\,\pi,\,0$, respectively. The vertical lines show the average arrival time. In all panels, the dark cyan dot-dashed lines show the semiclassical approximation, the green dashed lines show the standard method and the black solid lines show the quantum flux approach. }
    \label{ATDH_1D_AB}
\end{figure}
\begin{figure*}
    \centering
    \includegraphics[width = 15cm]{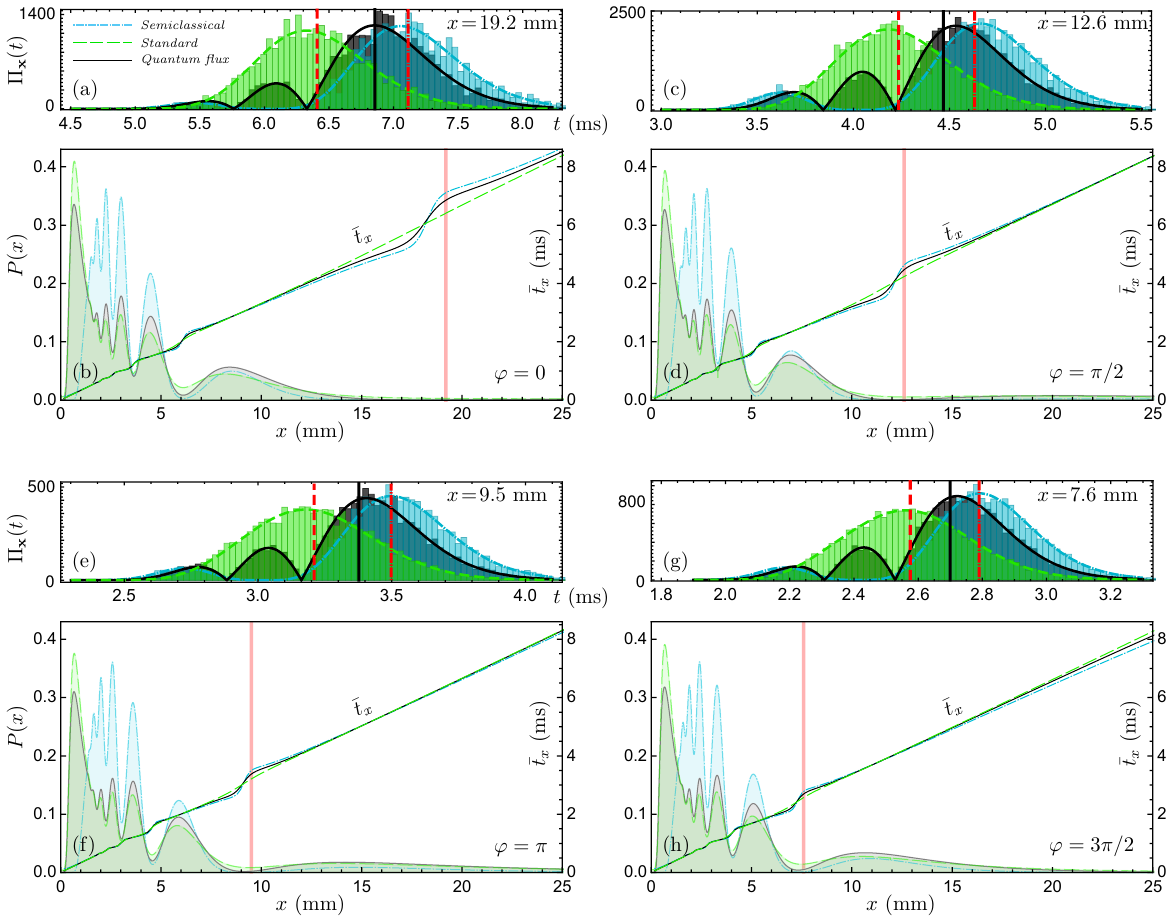}
    \caption{\textbf{Spatiotemporal arrival statistics for the double-slit experiment with a horizontal screen.} Filled curves in the panels (b), (d), (f) and (h) show the cumulative position probability density for the double-slit experiment with a horizontal screen placed at $L_y\!=\!15 \, \mu$m. The internal graphs in these panels represent the average time of arrival at each point of the screen, $\bar t _{\vb x}$. The panels (a), (c), (e), and (g) show the histograms of the local arrival time probability densities, $\Pi_{\vb x} (t|\vb x \!\in \! \mathbb{S})$, at the points $x=19.2$ mm for $\varphi=0$, $x=12.6$ mm for $\varphi=0.5\,\pi$, $x=9.5$ mm for $\varphi=\pi$, and $x=7.7$ mm for $\varphi=1.5\,\pi$ on the screen, respectively, which is marked with red lines in their lower panels. The histograms are associated with probability densities of local distributions, which are generated by numerical random sampling. In all panels, the dark cyan dot-dashed lines show the semiclassical approximation, the green dashed lines show the standard method and the black solid lines show the quantum flux approach. }
    \label{AATH_AB}
  \end{figure*}
The first quantity that we can study is the arrival time distribution on the horizontal screen. As one can see in Fig.\,\ref{ATDH_1D_AB}, by changing the relative phase, the patterns of the distributions and consequently average arrival times are shifted. However, the differences between the distributions and averages do not change much. In Fig.\,\ref{AATH_AB}, the arrival position distribution and the average arrival times at each point of the screen are shown. As in the previous figure, changing the relative phase causes a shift in the patterns of these quantities. Using this pattern shift, one can suggest an operational domain for experimental investigations. As the relative phase changes, we can adjust the position of the deviation regions of the average arrival times on the different arrival position distributions (see the solid vertical red lines in Fig.\,\ref{AATH_AB}). 
\begin{figure*}
    \centering
    \includegraphics[width = 15cm]{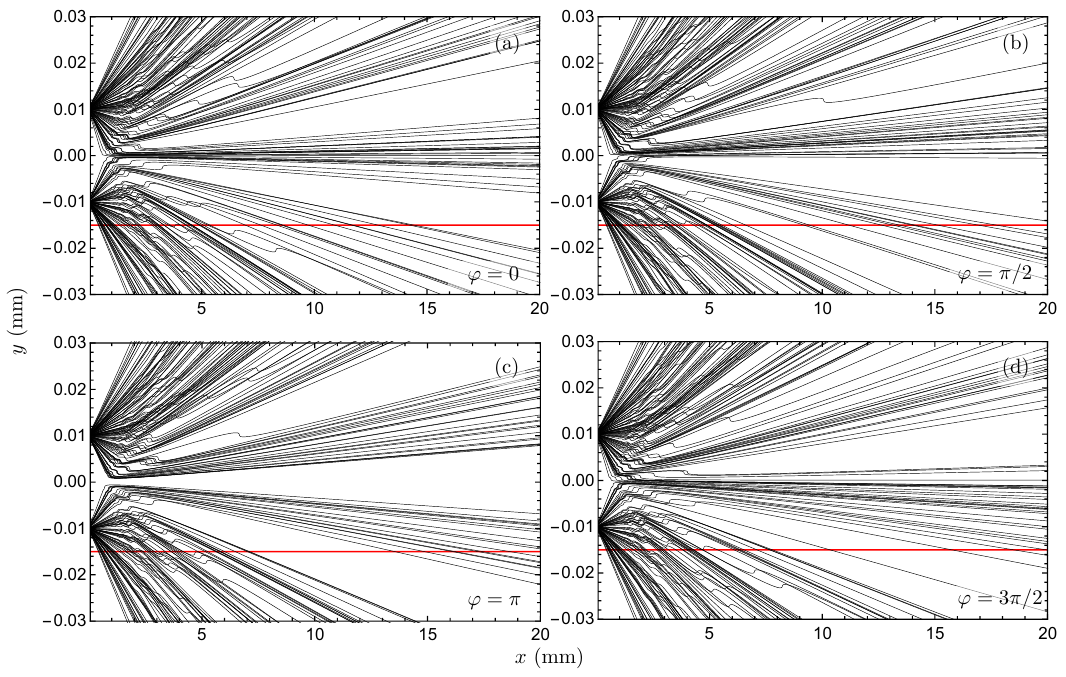}
    \caption{\textbf{Bohmian trajectories of particles that cross through the two-slit for four different relative phases.} The initial conditions are the same in all panels and 200 trajectories are computed in each one. Panels (a), (b), (c), and (d) correspond to the relative phases $\phi=0,\,0.5\,\pi,\,\pi,\,1.5\,\pi$, respectively.}
    \label{Traj_AB}
\end{figure*}
 \begin{figure*}
    \centering
    \includegraphics[width = 15cm]{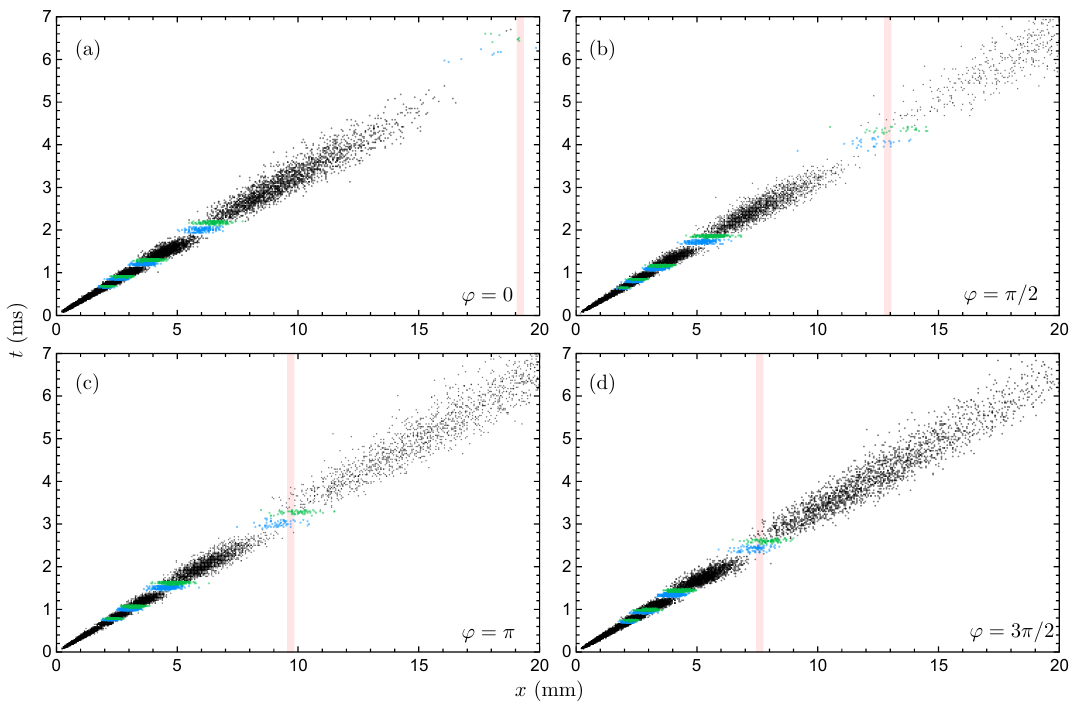}
    \caption{\textbf{Scatter plots of particles that arrive at the horizontal screen for four different relative phases.} The initial conditions are the same in all panels and Each of the scatter plots is generated using  $2\times 10^4$ Bohmian trajectories. Panels (a), (b), (c), and (d) correspond to the relative phases $\phi=0,\,0.5\,\pi,\,\pi,\,1.5\,\pi$, respectively. The vertical red lines correspond to the ones in Supplementary Figure \ref{AATH_AB}.}
    \label{SP_AB}
\end{figure*}

In fact, there is a trade-off between the number of particles that arrive and the scale of discrepancies among the predicted quantities. To compare the effects of the different phases more precisely, we have plotted in each panel of Fig.,\ref{AATH_AB} the histograms of local arrival distributions at the specified points, i.e., where the differences between averages are maximum. These points are highlighted by the vertical red lines in Fig.\,\ref{AATH_AB} (the deviation regions). In this analysis, we fixed the total number of particles that pass through the two-slit at $10^8$, so the number of particles that arrive at the deviation regions varies: about $10^4$ particles around $x=19.2$ mm for $\phi=0$, $2\times 10^4$ particles around $x=12.6$ mm for $\phi=0.5\,\pi$, $3\times 10^4$ particles around $x=9.5$ mm for $\phi=\pi$, and $5\times 10^4$ particles around $x=7.7$ mm for $\phi=1.5\,\pi$. These deviation regions are centered on the specified points and their thickness is assumed to be $0.2$ mm. The histograms in Fig.\,\ref{AATH_AB} show that the best fit between the predicted distributions and numerical experiments occurs around $x=7.7$ mm, and the worst one belongs to the distribution around $x=19.2$ mm. However, the later region has the biggest discrepancy between the average arrival times and the former region has the smallest one.
  
Fig.\,\ref{Traj_AB} depicts the effect of different relative phases on Bohmian trajectories. Changing the phase causes a shift in particle trajectories. It is worth noting that the ensemble of trajectories can be experimentally reconstructed using weak measurement techniques \cite{Sacha2011Observing,schleich2013reconstruction}, which can be compared with arrival distributions \cite{coffey2011comment,coffey2011reconstruction}. Fig.\,\ref{SP_AB} shows the scatter plots corresponding to the arrival times and positions of particles that arrive at the horizontal screen (horizontal red lines in Fig.\,\ref{Traj_AB}). From these plots, one can infer that the deviation regions of average arrival times (i.e., the solid vertical red lines in Figs.\,\ref{AATH_AB} and \ref{SP_AB}) correspond to the regions where Bohmian multi-crossing occurs. A thin laser sheet, as a detection screen, seems suitable to observe arrival distributions in these regions \cite{roncallo2023does,ott2016single,goussev2015manipulating}.

\section{Leavens-Muga debates and Galapon unitary collapse interpretation}\label{appendixB}
In \cite{leavens2002standard}, Leavens discusses some paradoxical behaviors that arise from Kijowski's arrival time distribution. This topic has also been commented on by other authors \cite{leavens2002standard,egusquiza2003comment,leavens2005reply,sombillo2016particle,das2021times}. One of the Leavens' comments is that Kijowski's distribution predicts non-zero arrival probabilities in some physically forbidden regions. For example, even if the wave function vanishes at a given point for all times, the Kijowski's arrival time distribution may leads to a non-zero probability for that point \cite{sombillo2016particle};
A vanishing wave function at a point implies the non-appearance of the particle at that point, and therefore it is expected that the arrival time distribution presents the same behavior as well.
In our double-slit setup, this situation occurs when the horizontal screen is placed at $L_y=0$ and the relative phase is set to $\varphi=\pi$. In this case, the position probability density, $|\psi|^2$, is zero on the screen for all times, while Kijowski's arrival time distribution is positive. Note that, in this case, the Bohmian arrival time distribution behaves as expected (see Fig.\ref{Traj_AB}). 
  \begin{figure}[t!]
    \centering
    \includegraphics[width = 8.6cm]{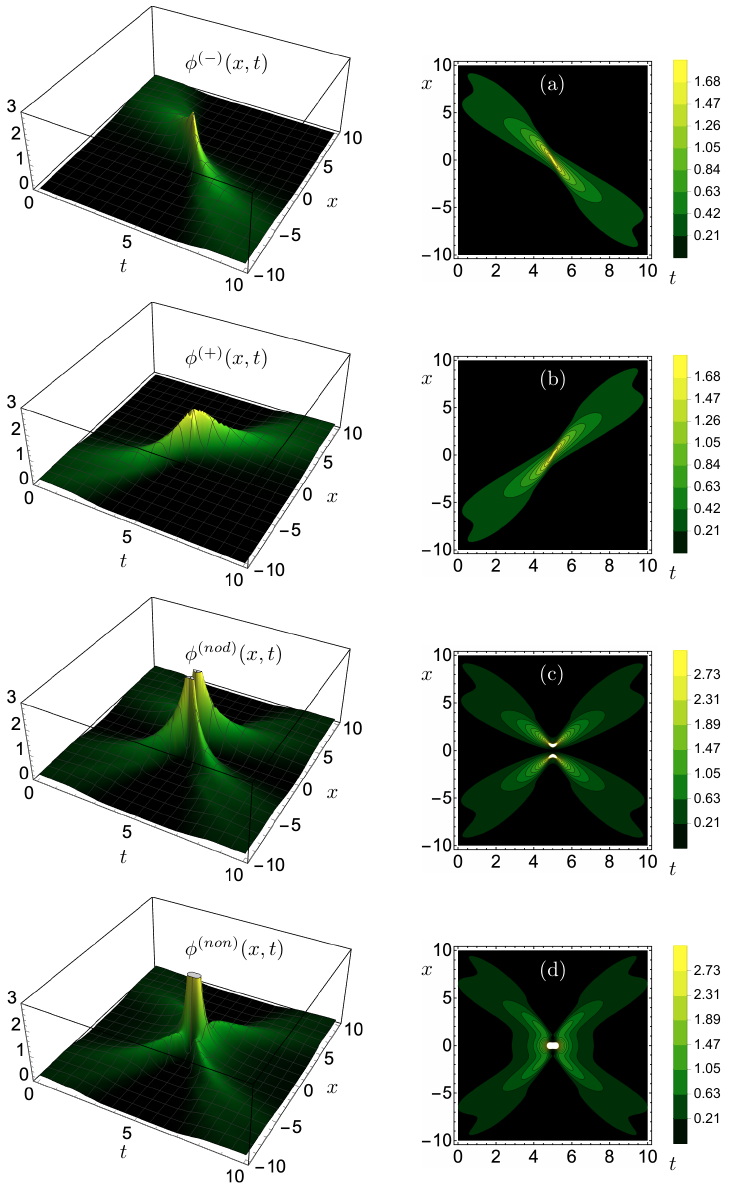}
    \caption{\textbf{Regularized free evolution of the eigenfunctions of the TOA operator for the left-right basis and the even-odd basis.} For all cases, the arrival point is set to the origin and the eigenvalue of time is $\tau=5$. The regularization parameter is $\epsilon = 0.1$ and all the other constants are $\hbar=m=1$. Panels (a), (b), (c), and (d) correspond to the left, right, odd, and even basis. }
    \label{time-eigen-function}
\end{figure}

In \cite{sombillo2016particle}, Galapon and Sombillo argued that the standard approach can avoid this problem if the eigen-functions of the arrival time operator are properly interpreted. To understand their analysis, we need to consider the unitary time evolution of the eigenfunctions of the time operator. The degenerate eigenfunctions of the arrival time operator $\hat{t}_L$ with eigenvalue $\tau$ can be expressed in terms of the following right-left basis in momentum representation \cite{muga1998space}:
$$
\tilde{\phi}^{(\pm)}_\tau(p)=\langle p|\tau^{(\pm)}\rangle=\sqrt{\frac{|p|}{m}}\frac{e^{-ipL/\hbar}}{\sqrt{2\pi\hbar}}e^{ip^2\tau/2m\hbar}\Theta(p),
$$
where $L$ is the arrival point. To obtain the corresponding amplitude in position space, $\langle x|e^{-i\hat{H}t}|\tau^{(\pm)} \rangle$, we need to integrate this in momentum space. However, the integral is divergent, so we introduce a converging factor $e^{-\epsilon p^2}$ such that \cite{sombillo2016particle}:
$$\phi^{(\pm)}(x,t)=\lim_{\epsilon\to 0}\int \tilde{\phi}^{(\pm)}(p) e^{ipx/\hbar-ip^2t/2m\hbar} e^{-\epsilon p^2} dp.$$
This time evolution of the arrival time eigen-functions exhibits an interesting behavior; when $t=\tau$, the evolving eigenfunction becomes a function with a singular support at the arrival point $L$ (i.e. the integral becomes divergent as $\epsilon\to 0$, see panels (a) and (b) of Fig.\,\ref{time-eigen-function}). In the light of this observation, Galapon proposed that the instant when the wave function has its minimum position uncertainty is interpreted as the particle's appearance \cite{galapon2009theory,sombillo2016particle}.
This mechanism suggests that a quantum particle can be materialized via smooth unitary evolution instead of abrupt collapse upon measurement. 
\begin{figure}[t]
    \centering
    \includegraphics[width = 8.6cm]{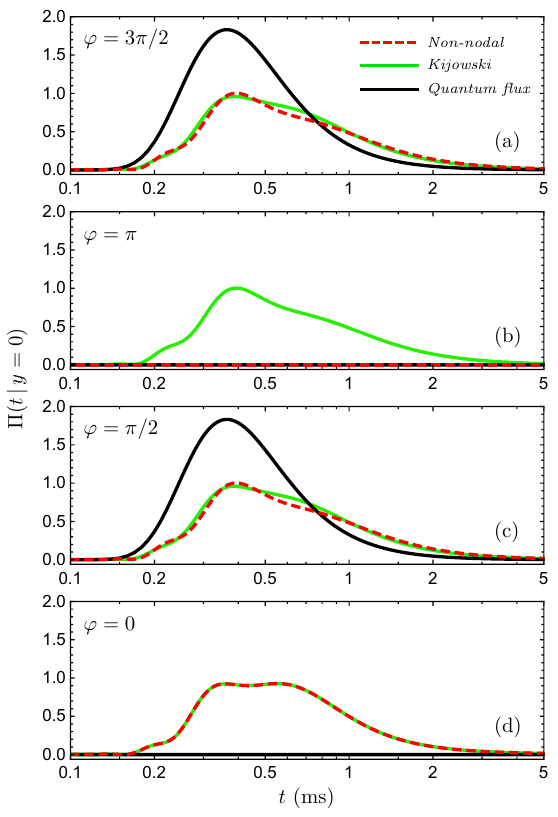}
    \caption{\textbf{Arrival time distributions of particles that arrive at the horizontal screen located at $L_y=0$ for four different relative phases.} Panels (a), (b), (c), and (d) show the arrival distribution, $\Pi(t|y=L_y)$, at screen distances $L_y\!=\!0\,\mu$m for four different relative phases $\phi=1.5\,\pi,\,\pi,\,0.5\,\pi,\,0$, respectively. In all panels, the red dashed lines show the non-nodal results, the green solid lines show the Kijowski method predictions, and the black solid lines show the quantum flux approach.
}
    \label{ATDH_1D_DA}
\end{figure}
In the conventional description, the initial wave function evolves unitarily and then collapses at the moment of detection, while in Galapon's unitary collapse approach \cite{galapon2009theory}, the order is reversed: The initial wave function of a particle collapses into one of the eigenfunctions of the arrival time operator right after the preparation and then the eigenfunction evolves naturally according to the Schrödinger equation into a function with singular support at the position of the detector.
It should be noted that the right-left eigenfunctions, $|\tau^{(\pm)}\rangle$, are not the only basis for the arrival time operator. A superposition of these eigenfunctions leads to the even-odd parity basis, $(|\tau^{(+)}\rangle\pm|\tau^{(-)}\rangle)/\sqrt{2}$, which in momentum representation are given by \cite{sombillo2016particle}:
\begin{eqnarray}
\tilde{\phi}^{(odd)}_\tau &=&\frac{\phi^{(+)}_\tau+\phi^{(-)}_\tau}{\sqrt{2}}=\sqrt{\frac{|p|}{m}}\frac{e^{-ipX/\hbar}}{\sqrt{2\pi\hbar}}e^{ip^2\tau/2m\hbar}\text{sign}(p)\nonumber\\
\tilde{\phi}^{(even)}_\tau&=&\frac{\phi^{(+)}_\tau-\phi^{(-)}_\tau}{\sqrt{2}}=\sqrt{\frac{|p|}{m}}\frac{e^{-ipX/\hbar}}{\sqrt{2\pi\hbar}}e^{ip^2\tau/2m\hbar}\nonumber
\end{eqnarray}\label{nod-non}
The eigenfunctions with even parity are also called non-nodal while those with odd parity are referred to as nodal. The non-nodal eigenfunctions exhibit the same behavior as that of the left-right eigen-functions, where they become a function with a singular support at the arrival point at a time equal to the arrival time operator eigenvalue. Such unitary collapse can be interpreted as particle appearance at the arrival point. However, the nodal eigen-function is zero at the arrival point for all times $t$ despite having a minimum position uncertainty at $t=\tau$ (see panels (c) and (d) of Fig.\,\ref{time-eigen-function}).
Galapon and Sombillo argue that if the support of the initial wave function be include positive and negative momentum, then the state can collapse into either the non-nodal or the nodal  eigen-functions of arrival time operator. Thus, if one is only interested in arrival time measurement with particle detection (appearance), then the relevant arrival time distribution come just from the non-nodal eigen-functions, i.e. $\Pi^{(non)}\equiv|\langle \tau^{(non)}|\psi_0\rangle|^2$. In Fig.\,\ref{ATDH_1D_DA}, the non-nodal arrival time distribution, $\Pi^{(non)}$, is compared with Kijowski's arrival time distributions and quantum flux, for a double-slit setup with a horizontal screen placed at $L_y=0$, and for different relative phases. As it is shown in figure \ref{ATDH_1D_DA}, the non-nodal arrival time distribution, like the quantum flux arrival time distribution, vanishes when $\varphi=\pi$, as expected, while Kijowski's arrival time distribution does not show this behavior. This fact is a very clear contrast between these proposals, which can be directly tested by the proposed double-slit setup.

\typeout{}
\bibliography{bibliography}

\end{document}